%% file: main.tex
\documentclass[12pt, titlepage]{report}

\usepackage[english]{babel}
\usepackage[utf8x]{inputenc}
\usepackage[T1]{fontenc}
\usepackage{listings}
\usepackage{color}
\usepackage{float}
\usepackage{enumitem}
 
\definecolor{codegreen}{rgb}{0,0.6,0}
\definecolor{codegray}{rgb}{0.5,0.5,0.5}
\definecolor{codepurple}{rgb}{0.58,0,0.82}
\definecolor{backcolour}{rgb}{0.95,0.95,0.92}
\lstdefinestyle{mystyle}{
    backgroundcolor=\color{backcolour},   
    commentstyle=\color{codegreen},
    keywordstyle=\color{magenta},
    numberstyle=\tiny\color{codegray},
    stringstyle=\color{codepurple},
    basicstyle=\footnotesize,
    breakatwhitespace=false,         
    breaklines=true,                 
    captionpos=b,                    
    keepspaces=true,                 
    numbers=left,                    
    numbersep=5pt,                  
    showspaces=false,                
    showstringspaces=false,
    showtabs=false,                  
    tabsize=2
}
 
\makeatletter
\def\@makechapterhead#1{%
  {\parindent \z@ \raggedright \normalfont
    \interlinepenalty\@M
    \Huge \bfseries \thechapter\ \ |\ \ #1\par\nobreak
    \vskip 40\p@
  }}
\makeatother

\usepackage[a4paper,top=3cm,bottom=2cm,left=3cm,right=3cm,marginparwidth=1.75cm]{geometry}

\usepackage{amsmath}
\usepackage{graphicx}
\usepackage[colorinlistoftodos, textsize=tiny]{todonotes}
\usepackage[colorlinks=true, allcolors=black]{hyperref}
\usepackage{caption}
\usepackage{subcaption}
\usepackage{scrextend}

\title{Investigating the OPS intermediate representation to target GPUs in the Devito DSL}
\author{Vincenzo Pandolfo}
\begin{document}
\input{title/title.tex}
\parskip 1.5ex % paragraph spacing
\renewcommand{\baselinestretch}{1.5} % line spacing

%\thispagestyle{empty}
%\clearpage\mbox{}\clearpage

\begin{abstract}
The Devito DSL is a code generation tool for the solution of partial differential equations using the finite difference method specifically aimed at seismic inversion problems.

In this work we investigate the integration of OPS, an API to generate highly optimized code for applications running on structured meshes targeting various platforms, within Devito as a mean of bringing it to the GPU realm by providing an implementation of a OPS backend in Devito, obtaining considerable speed ups compared to the core Devito backend.
\end{abstract}

%\thispagestyle{empty}
%\clearpage\mbox{}\clearpage

\renewcommand{\abstractname}{Acknowledgements}
\begin{abstract}
I would like to express my gratitude to my supervisors, Prof. Paul Kelly for being an invaluable source of wisdom and knowledge throughout my university career as my academic tutor and a fantastic guide during the course of this project, and Dr. Fabio Luporini for being always available to help and point me in the right direction.

Many thanks to Nicolai Stawinoga, for allowing me to use his machine for the performance evaluations presented in this thesis and to the Devito and OPS developers who where always ready to answer my questions.

Finally, I would like to thank my girlfriend, Cecilia, for her continued support and my family, without whom I wouldn't be here.
\end{abstract}

%\thispagestyle{empty}
%\clearpage\mbox{}\clearpage

\tableofcontents
% \listoffigures
% \listoftables

\input{introduction/introduction.tex}
\input{background/background.tex}
\input{project/project.tex}
\input{evaluation/evaluation.tex}
\input{conclusion/conclusion.tex}

\bibliographystyle{ieeetr}
\bibliography{main}

\end{document}

%% file: title/title.tex
\begin{titlepage}

\newcommand{\HRule}{\rule{\linewidth}{0.5mm}} % Defines a new command for the horizontal lines, change thickness here

%----------------------------------------------------------------------------------------
%	LOGO SECTION
%----------------------------------------------------------------------------------------

\includegraphics[width=8cm]{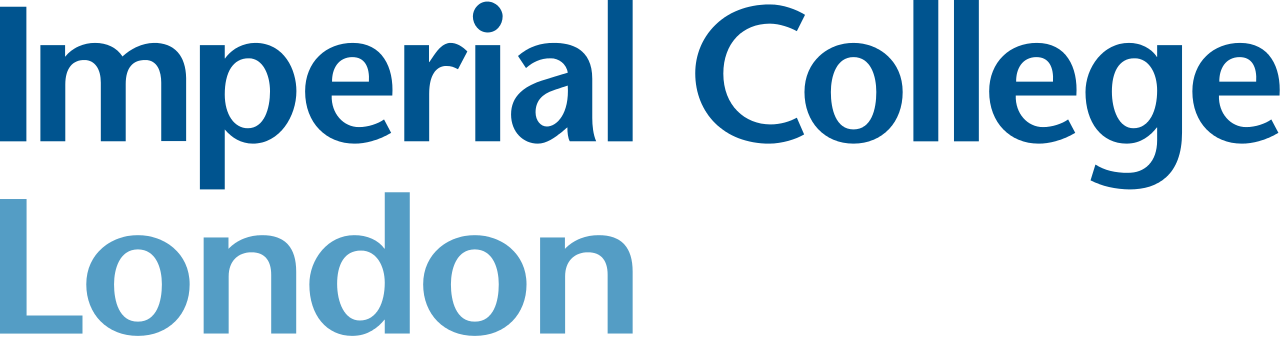}\\[1cm] % Include a department/university logo - this will require the graphicx package
 
%----------------------------------------------------------------------------------------

\center % Center everything on the page

%----------------------------------------------------------------------------------------
%	HEADING SECTIONS
%----------------------------------------------------------------------------------------

\textsc{\LARGE MEng Individual Project}\\[1.5cm] % Name of your university/college
\textsc{\Large Imperial College London}\\[0.5cm] % Major heading such as course name
\textsc{\large Department of Computing}\\[0.5cm] % Minor heading such as course title

%----------------------------------------------------------------------------------------
%	TITLE SECTION
%----------------------------------------------------------------------------------------
\makeatletter
\HRule \\[0.4cm]
{ \huge \bfseries \@title}\\[0.4cm] % Title of your document
\HRule \\[1.5cm]
 
%----------------------------------------------------------------------------------------
%	AUTHOR SECTION
%----------------------------------------------------------------------------------------

\begin{minipage}{0.4\textwidth}
\begin{flushleft} \large
\emph{Author:}\\
\@author % Your name
\end{flushleft}
\end{minipage}
~
\begin{minipage}{0.4\textwidth}
\begin{flushright} \large
\emph{Supervisors:} \\
Prof. Paul Kelly \\%[1.2em] % Supervisor's Name
% \emph{Second Marker:} \\
Dr. Fabio Luporini % second marker's name
\end{flushright}
\end{minipage}\\[2cm]
\makeatother

% If you don't want a supervisor, uncomment the two lines below and remove the section above
%\Large \emph{Author:}\\
%John \textsc{Smith}\\[3cm] % Your name

%----------------------------------------------------------------------------------------
%	DATE SECTION
%----------------------------------------------------------------------------------------

{\large June 17, 2019}\\[2cm] % Date, change the \today to a set date if you want to be precise

\vfill % Fill the rest of the page with whitespace

\end{titlepage}

%% file: introduction/introduction.tex
\chapter{Introduction}

Numerical solution of partial differential equations (PDEs) has many different real world applications. It is a computationally intensive problem and therefore has been the focus of decades of research, with researchers trying to optimize different stages of the computation: from the formulation of the PDE itself to more low level optimizations.

Historically, this would be done with bespoke software for specific applications, making it very difficult to develop and test new algorithms and formulations, to port existing software to new architectures that require different optimizations, and to apply new optimizations to a wide range of different problems.

To overcome this issue, a possible solution is to provide general purpose tools that can, from a high level specification of a PDE, generate high performance, parallel code that can run on a plethora of different architectures. Such tools would allow researchers to try out new ideas on large data sets quickly, without having to worry about the optimization side of the software that would otherwise be required.

Domain Specific Languages (DSLs) are one such tool that has been proven successful in various fields. They provide specialists with tools to formulate their ideas (algorithms, PDEs, etc.) in ways that are familiar to them and relevant to their domain, abstracting away the complexity of optimizing the code that is actually executed.
Successful examples of DSLs can be found in the finite element space in tools such as Firedrake \cite{Rathgeber2016} or FEniCS \cite{AlnaesBlechta2015a}, both making use of the Unified Form Language \cite{AlnaesEtAl2012} developed for the latter.

The focus of this project is a DSL and code generation tool called Devito \cite{devito-api} whose goal is to generate highly optimised code for the solution of PDEs using the finite difference method. In particular, Devito aims to allow users to efficiently work on seismic inversion problems. One major concern in Devito is having to incorporate specific optimizations for every targeted platform.

To alleviate these problems part of the optimization done by Devito can be offloaded to existing tools that can efficiently optimise parts of the code that would be generated by Devito. One such tool that is currently being used is YASK \cite{Yount2016YASKYetAS}, a tool to optimize stencil code targeting Intel CPUs.

OPS \cite{Reguly:2014:ODS:2691166.2691173} is an API to generate highly optimized code for applications running on structured meshes, targeting CUDA, OpenCL and MPI among others. The OPS representation lends itself well to Devito, as it can be used to optimize big parts of the generated code, potentially giving performance benefits to already targeted architectures and most importantly allowing it to target multiple architectures effectively, reducing the workload for Devito developers.

This project explores the possibility of integrating OPS within Devito as a compiler backend. This will ultimately help to inform a decision on whether adopting OPS as a backend library is a valuable path forward for Devito, especially in terms of expanding in the GPU realm that is becoming increasingly popular in the HPC world.

This thesis makes the following contributions:
\begin{itemize}
    \item Generation of C code that uses the OPS API within Devito
    \begin{itemize}
        \item Providing the bare minimum to target any platform supported by OPS
    \end{itemize}
    \item Compilation and execution of the generated code from Devito
    \begin{itemize}
        \item Provides ways to target different platforms (CUDA, OpenMP, MPI) using OPS libraries via Devito, easily extendable to target more.
    \end{itemize}
    \item Evaluation
    \begin{itemize}
        \item  To demonstrate the viability of using OPS in Devito, benchmarks for different targeted platforms have been run and their results evaluated. We also evaluate the software quality of the OPS library as this also represents a deciding factor regarding its adoption
    \end{itemize}
    \item Recommendations on what steps forward Devito should take to target GPUs
\end{itemize}

At the end of this thesis the reader will have a clear idea of what the effort required to implement a OPS backend is, what performance to expect and what some ways to expand this work to allow Devito to better target new platforms are.

%% file: background/background.tex
\chapter{Background and related work}

This chapter aims to provide an understanding of the principal components of this project (Devito and OPS) and explore related work.

The first section will give a top level overview of Devito, including some more relevant low level details.

In the second section OPS will be presented, giving a brief description of the API and the features it provides.

Related work will then be discussed, in particular the YASK backend in Devito.

\section{Devito}
Devito is a DSL and code generation tool for the solution of PDEs using the finite difference method. In particular it is focused on the generation of highly efficient kernels for the solution of seismic inversion problems. It uses SymPy \cite{10.7717/peerj-cs.103} to define high level operators from symbolic equations that are then converted into C code that can run efficiently on a specific target architecture.

The use of symbolic equations allows Devito to be extremely user-friendly to domain specialists, enabling them to define complex operators in little code, hiding from them the complexity of code optimisation. This makes it extremely quick for users to develop new solvers (or rewrite existing ones) that can target different architectures.

\subsection{Top level overview}

The main building blocks of Devito programs are \texttt{Operator}s that contain the problem specification: the data they operate on and the expressions that will be evaluated over this data. Once a user has defined their \texttt{Operator}, Devito will use it to generate optimised stencil code that is run in (optimised) loops to apply the equations over the input data.

The Devito CFD tutorial \cite{devito-cfd-tutorial} provides us with a simple example of an Operator for a 2D linear convection:

\begin{lstlisting}[language=Python]
from devito import Eq, Grid, Operator, TimeFunction

grid = Grid(shape=(nx, ny), extent=(2., 2.))
u = TimeFunction(name='u', grid=grid)

# Initialize the input data
init_hat(field=u.data[0], dx=dx, dy=dy, value=2.)

# Specify the `interior` flag so that the stencil is only
# applied to the interior of the domain.
eq = Eq(u.dt + c*u.dxl + c*u.dyl, subdomain=grid.interior)

stencil = solve(eq, u.forward)

# Create an operator that updates the forward stencil point
op = Operator(Eq(u.forward, stencil, subdomain=grid.interior))

# Apply the operator for a number of timesteps
op(time=nt, dt=dt)
\end{lstlisting}

Functions are identified by a name (in the example, \texttt{u}) and operate on a computational \texttt{Grid}.
\texttt{Function}s are abstractions describing spatially varying functions. An extension of these are \texttt{TimeFunction}s: these describe spatially varying time dependent discrete functions.

\texttt{SparseFunction} and \texttt{SparseTimeFunction} are sparse functions that operate only on a subset of the grid. 

\texttt{u.forward}, \texttt{u.dx1} and \texttt{u.dy1} are examples of useful finite difference shortcuts provided by Devito in order to aid with the formulation of FD stencils.

\begin{figure}[H]
    \centering
    \includegraphics[width=\columnwidth]{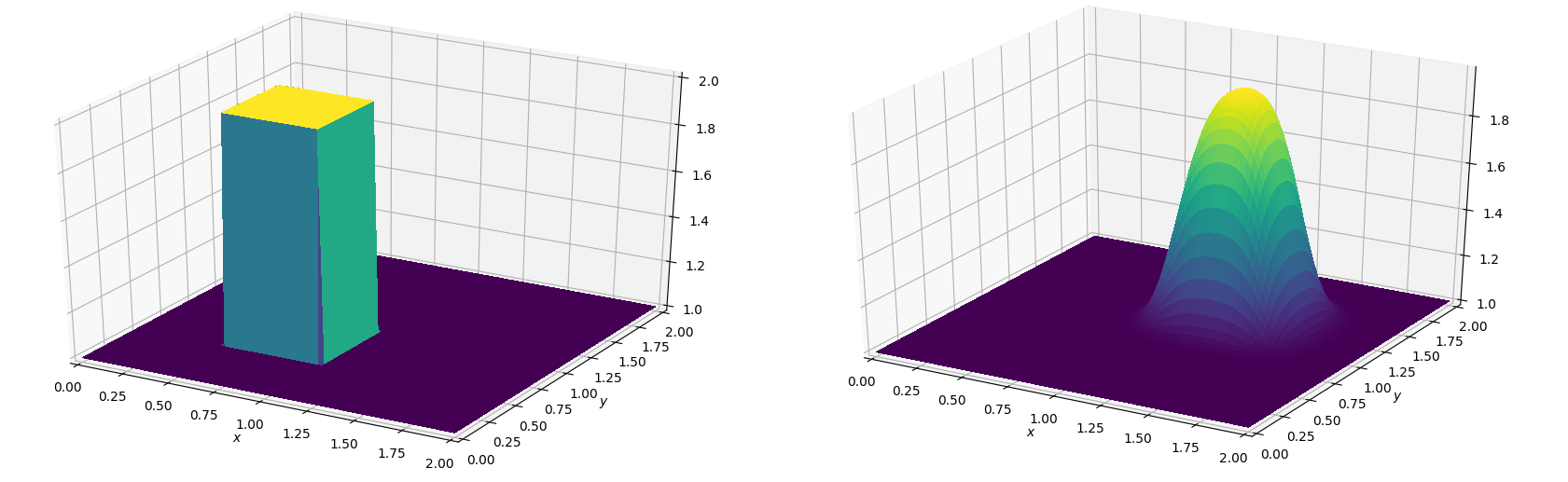}
    \caption{The input data (left) and the result (right) of the execution of the example \texttt{Operator} in section 2.1.1 \cite{devito-cfd-tutorial}}
    \label{fig:convection_result}
\end{figure}

\subsection{Optimizations}
Once Operators are defined by the user, they undergo a process of optimizations in a sequence of steps, most notably executed by the Devito Symbolic Engine (DSE) and the Devito Loop Engine (DLE), which will be covered in more detail shortly \cite{DBLP:journals/corr/abs-1807-03032}.

The goal of this project is to offload part of the optimization process to OPS (described in more detail later), in order to allow Devito to target GPUs without having to write the required optimizations directly. The nature of the Devito optimisation pipeline makes it possible to swap out parts of it with different components, so that external tooling can be used at the right stage when appropriate.

\subsubsection{Devito Symbolic Engine}

The DSE is tasked with reducing the operational count of groups of equations sharing the same domain (called \texttt{Cluster}s) via means of symbolic manipulation. This stage does not directly impact the nature of the loop nests generated by Devito, but it is focused on the optimization of the expressions themselves with methods such as common sub-expressions elimination, FD weights factorization and alias detection.

Once this step is performed, this intermediate representation is transformed into an Iteration/Expression Tree (IET), an abstract syntax tree that represents the structure of the loop nests.

\subsubsection{IET and Devito Loop Engine}

When the IET is built, equations are embedded in \texttt{Expression}s, nested inside \texttt{Iteration}s representing the loop nests. The IET is then used for further optimizations performed by the DLE. These include SIMD vectorization, loop blocking and parallelism via OpenMP.

Different backends can provide alternative loop optimization engines. One example of such a back end is the YASK Loop Engine (YLE) that transforms suitable elements of the IET into a format that YASK can process, therefore offloading most of the optimization work to it.

The OPS backend discussed in this report is similar in nature to the YASK one, as IETs are transformed in such a way that the generated code uses the OPS API, leaving the final optimization tasks to the OPS compiler.

\subsection{Compilation and execution}

Once all these steps are performed, variable declarations, headers and profiling instrumentation is inserted in the IET. Once this is done, a IET visitor generates a \texttt{CGen}\cite{cgen} AST that is then transformed into a source code file, JIT-compiled and loaded into the \textit{Python} environment. The code can then be executed directly from an \texttt{Operator} via a function used as a default entry point.

\begin{figure}[h]
    \centering
    \includegraphics[width=\linewidth]{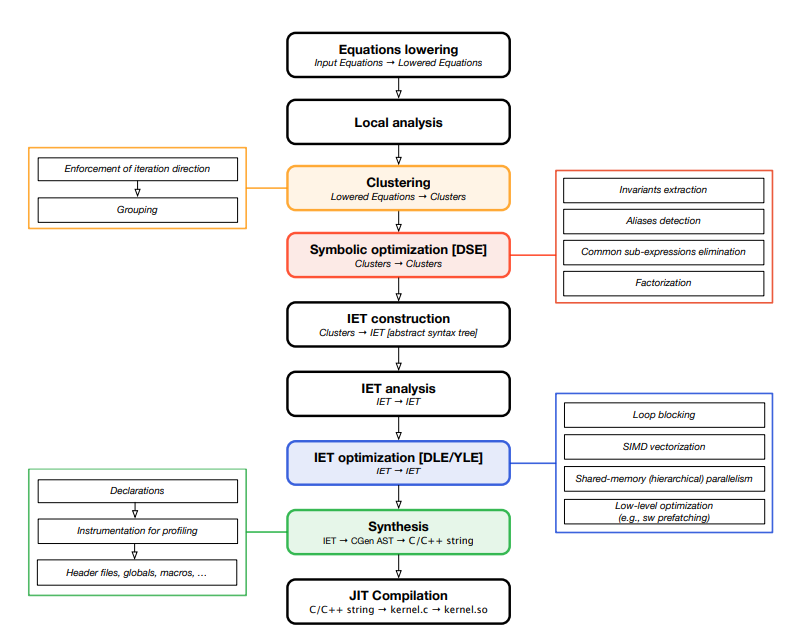}
    \caption{The Devito pipeline \cite{DBLP:journals/corr/abs-1807-03032}}
    \label{fig:devito_pipeline}
\end{figure}
\newpage
\subsection{IET nodes}

Most of the focus of this report is on manipulation of IETs and therefore of IET nodes. We will now give a brief overview of some IET nodes that will be used in this report.
\begin{itemize}
    \item \textbf{Expression}: encapsulates a \texttt{ClusterizedEq}, a Sympy equation with associated iteration and data space. It is rendered in code generation as an assignment
    \item \textbf{Call}: represents a function call
    \item \textbf{Callable}: represents a function definition, including its signature and body
    \item \textbf{Iteration}: a for loop
    \item \textbf{IterationTree}: represents a series of nested iterations
    \item \textbf{Element}: general purpose node that can contain different \texttt{cgen} objects
\end{itemize}

\section{OPS}
The Oxford Parallel library for Structured mesh solvers (OPS)\cite{Reguly:2014:ODS:2691166.2691173} is a Domain Specific Active Library for the development of applications operating on multi-block structured meshes. It provides an abstraction and API to enable automatic parallelization and optimization of computations on multi-block grid computations. The OPS abstraction is composed of four main components:
\begin{itemize}
    \item Blocks: high-level definition of the grid, with dimensionality but no size
    \item Dataset: the actual data contained in blocks
    \item Halos: interface between datasets on different blocks
    \item Computations: operations to apply to grid points
\end{itemize}

The API is straightforward, as we will see:
\begin{itemize}[listparindent=\parindent]
    \item \texttt{ops\_block ops\_decl\_block(num\_dims, ...)} defines a block with \texttt{num\_dims} dimensions
    \item \texttt{ops\_dat ops\_decl\_dat(block, size[], ...)} defines a dataset on \texttt{block} with size \texttt{size2}. Ownerwship of the data will be passed to OPS and it will then only be accessible via \texttt{ops\_dat} handles
    \item \texttt{void ops\_par\_loop( void (*kernel)(...), block, ndim, range[], arg1, ..., argN)} defines a parallel loop over \texttt{block} with the given dimensions and range that applies \texttt{kernel} with the given arguments
    \item \texttt{ops\_arg ops\_arg\_dat(...)} defines an argument that can be used to pass a dataset to \texttt{ops\_par\_loop}
    \item \texttt{ops\_arg ops\_arg\_gbl(...)} defines an argument that can be used to pass a global scalar or small array to \texttt{ops\_par\_loop}
\end{itemize}

Once users have written their code using OPS, this will go through the OPS source-to-source translator, a Python library that will generate optimized code for the target architecture. To do so, it will parse calls to \texttt{ops\_par\_loop} in order to extract the required information and will transform it into very specific and optimized loop code.
This code will then be linked against OPS libraries (the OPS back end) in order to generate high performance executables.

This back-end gives important features to OPS, some examples being:

\textit{Distributed memory parallelism}: using MPI, OPS can automatically partition blocks and datasets over multiple processes

\textit{Checkpointing and recovery}: thanks to OPS taking ownership of the data, it can keep track of the transformations applied to the data in order to periodically store it on disk (\textit{checkpointing}) and automatically recover in case of failures.

\textit{Tiling}: Using runtime execution data OPS could be able to analyze data dependencies and generate effective tiling strategies.

These different optimizations can be applied by OPS on multiple architectures, with no work required by the user as they will only need to write their code using the OPS abstraction once. This characteristic makes it a great candidate as a Devito back-end as it sits on a much lower level. Source code generation in Devito can target OPS instead of a specific architecture, making Devito easier to reason about and freeing Devito developers from the burden of thinking of low level optimizations and allowing them to concentrate on more domain specific issues.

\section{Related work}

\subsection{YASK}

YASK (Yet Another Stencil Kernel) is a framework that allows the generation of high performance code targeted at Intel Xeon and Xeon Phi processors. It provides multiple optimizations, including:

\begin{itemize}
    \item Vector-folding
    \begin{itemize}
        \item Traditionally, multi-dimensional vectors are stored as a sequence along one dimension. This is inefficient for multi-dimensional stencils. Vector-folding is a data storage technique that minimizes memory accesses by storing multi-dimensional data in a vector \cite{7336272}. 
    \end{itemize}
    \item Multi-level parallelism via OpenMP
    \begin{itemize}
        \item Outer loops are parallelized across cache blocks size that are then ulteriorly split in inner loops. This allows an increase in temporal cache locality, avoiding eviction of data being used by concurrent threads.
    \end{itemize}
    \item Multi-socket and nodes processing via MPI
    \item Space tiling
    \begin{itemize}
        \item Also known as loop blocking, space tiling optimizes memory access by applying loop transformations to split the iteration space in chunks.
    \end{itemize}
    \item Time tiling
    \begin{itemize}
        \item Time tiling applies the idea of space tiling across time steps. YASK accomplishes this using a technique called \textit{temporal wave-front tiling} \cite{YOUNT2019903}.
    \end{itemize}
\end{itemize}

It provides an API in both C++ and Python.

\subsubsection{Devito YASK backend}

An ongoing project within Devito is integrating YASK as a backend, in a similar fashion to what this project has done with OPS. Currently, around 70\% of the Devito API is supported by the YASK backend. \cite{DBLP:journals/corr/abs-1807-03032}

\subsection{Loo.py}

Loo.py \cite{DBLP:journals/corr/Klockner14} is a Python library that provides a data model for the definition of array computation and a set of transformations that can be applied on these definitions to generate optimized CUDA/OpenCL code. The transformations to be applied are decided by the user and include loop tiling, vectorization, instruction-level parallelism among others.

Computations are defined using the \texttt{isl} library syntax \cite{isl} and then transformed by the user using the provided (or custom) transformations. This way, the user has complete control of what happens to their code, leaving the Loo.py only the task of applying the transformations and generating the code without trying to do anything clever to further optimize the generated code.

The resulting code can then be executed using \texttt{pyOpenCL}, or printed to be compiled and executed via other means if a different platform than OpenCL is being targeted (for example, CUDA).

\subsection{Mint}

Mint \cite{mint} is a source-to-source C translator that generates optimized CUDA code based on annotations (pragmas). Mint specifically targets stencil computation: by restricting the application space in which it operates it can perform more effective optimizations for that problem class.

Optimizations are performed depending on the results of analysis of the stencil done by a component called \textit{stencil analyzer}. The results are then passed to the \textit{On-chip Memory Optimizer} that uses this information to effectively optimize optimize memory accesses using registers and shared memory in order to avoid neighbouring points being loaded multiple times by independent.

Another optimization performed by Mint is \textit{Loop Aggregation} (or loop blocking) in which chunks of the iteration space are assigned to a single thread in order to minimize shared memory accesses. 

Mint has achieved good performance, with around 80\% of performance compared to hand-made optimized CUDA code.

\newpage
\subsection{Lift}

Lift \cite{lift} is a Scala library whose goal is achieving performance portability for DSL and library developers by providing a set of parallel primitives that can be composed to generate high performance stencil code for GPUs. 

Lift applications are expressed as a sequence of functional primitives and are optimized through a set of 1D rewrite rules. Users can compose these rewrite rules to express an optimize multidimensional stencil code. The technicalities of the optimizations that can be applied by Lift are hidden from the user who, however, has complete control over how high-level optimizations are applied.

The resulting stencil program is then converted in a sequence of low-level OpenCL-specific primitives that are then transformed in optmized OpenCL code.

Lift has achieved performance comparable to hand-optimized OpenCL code \cite{lift2}.

%% file: project/project.tex
\chapter{Code generation}

This chapter describes the process of generating OPS code within Devito.

First, a top level overview of the generation process will be provided. This will be expanded in further sections to explain how OPS kernels are:
\begin{itemize}
    \item Generated - this includes:
    \begin{itemize}
        \item Translation of Sympy expressions to use OPS accessors
        \item Creation of an header file to contain the kernels
    \end{itemize}
    \item Invoked - this includes:
    \begin{itemize}
        \item Declaration of datasets for use in OPS
        \item Invocation of kernels inside an OPS parallel loop
    \end{itemize}
\end{itemize}

When necessary, extensions to the core Devito functionality is described.

\section{Overview}

Devito allows the creation of different backends that can modify the default behaviour to various degrees. We utilise this feature to introduce an OPS backend whose goals are to generate, compiles and run code that uses the OPS API.

In order to do this a new \texttt{OperatorOPS} that subclasses Devito's default \texttt{Operator} has been created. This will be loaded as the \texttt{Operator} to be used automatically by Devito when the \texttt{ops} backend is selected (this is done by setting the \texttt{DEVITO\_BACKEND} environment variable).

All necessary operations (before execution) are performed as a sequence of method invocations within the \texttt{Operator} constructor. This allows subclasses to override only parts of the pipeline that are relevant to what needs to be achieved.

In the case of the \texttt{OperatorOPS} the main method to override is \texttt{\_finalize\_iet}: this method is called after all symbolic manipulation and IET construction (including loop generation) is completed and is tasked with specializing the IET depending on the backend being used. For the \texttt{core} backend (the default Devito backend) this means running the Devito Loop Engine (DLE). For our purposes this will mean transforming offloadable loop nests into OPS kernel, invoked through calls to \texttt{ops\_par\_loop} as described in the next sections.

\begin{figure}[h]
    \centering
    \includegraphics[width=\linewidth]{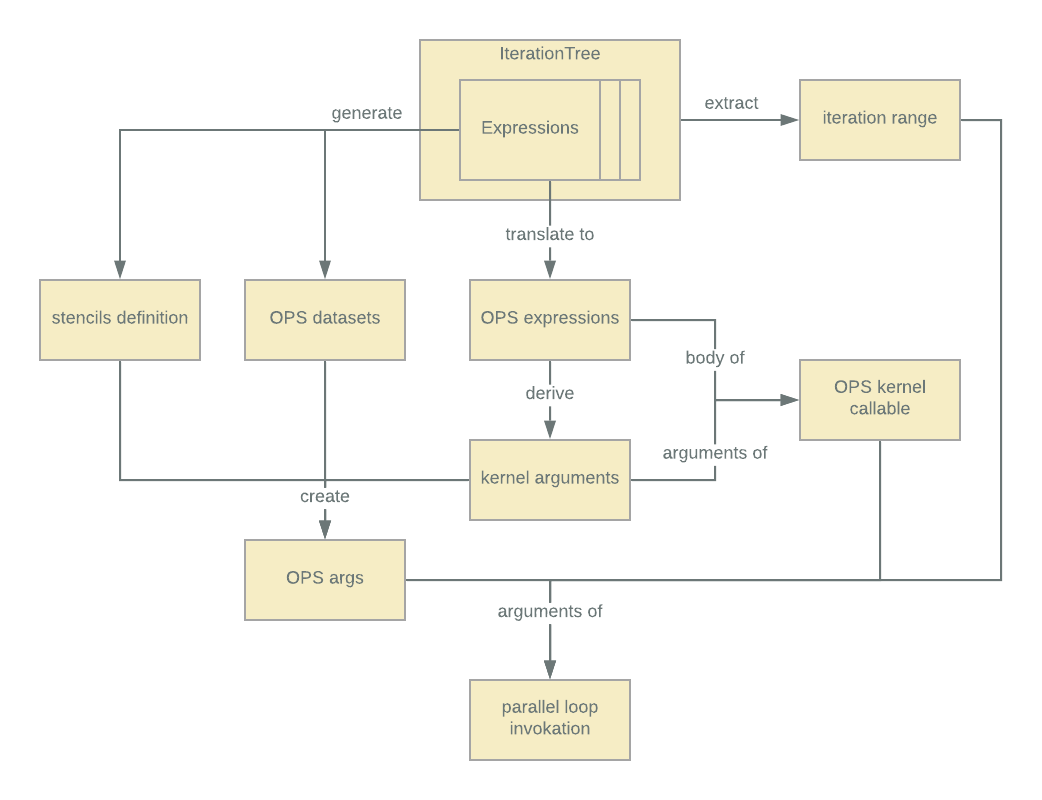}
    \caption{Code generation pipeline for the OPS backend}
    \label{fig:codegen}
\end{figure}

\subsubsection{Finding offloadable loops}

In order for a loop nest to be offloaded to OPS it has to be \textit{affine}, meaning that all of its array accesses are either constant or affine functions of the iteration variables. This is because non affine loops can lead to data races if parallelised and therefore can't be further optimized easily.

To find offloadable trees the \texttt{find\_affine\_trees} utility function (originally from the YASK backend that uses it for the very same purpose) is used. This returns a list of all offloadable  \texttt{IterationTree}s. We iterate on this list and for each such tree we generate:
\begin{itemize}
    \item An OPS kernel
    \item Dataset declarations
    \item Parallel loop information declaration (stencils, iteration range)
    \item Parallel loop invocation
\end{itemize}

The parallel loop invocation will substitute the \texttt{Section}s containing the offloaded \texttt{IterationTree}s with OPS loop invocations (therefore removing the actual loop from the IET) using a \texttt{Transformer} visitor: this visitor takes an IET and a map of substitutions and returns a new IET with these substitutions applied.

We then return a \texttt{List} node containing, in this order:
\begin{itemize}
    \item OPS initialization (a call to \texttt{ops\_init} and creation of the \texttt{ops\_block})
    \item Dataset declarations
    \item Parallel loop information declaration
    \item The newly created IET
    \item A call to \texttt{ops\_timing\_output} to print profiling information from OPS
    \item A call to \texttt{ops\_exit} to terminate OPS
\end{itemize}

We also include the headers for OPS and our generated kernels by adding them to the \texttt{\_includes} field in the operator.

\section{OPS Kernel creation}

When offloading loops to OPS the first step is to extract the \texttt{Expression}s representing the actual computation in a Kernel function that can be passed to OPS.

To do this the expressions have to be translated to use OPS accessor macros instead of loop indexes to index data within arrays, then placed in the body of a \texttt{Callable} (a Devito IET node representing a function definition) having as parameters the functions and constants used in the expressions. These are found using a \texttt{FindSymbols} visitor, already present in Devito.

\subsection{Expression translation}

An implementation of Devito expression translation to OPS expressions was already developed \cite{devito-ops-transl} and it remained mostly unchanged in logic, with some alterations to enable a clearer naming of indexed functions, facilitating the insertion of the correct \texttt{OPS\_ACC} accessor during a second pass, mark arrays as either read or write and accessing constants by dereferencing their pointers (since everything must be passed into an OPS Kernel as a pointer).

In order to translate them, expressions are recursively visited and rebuilt.
Changes are applied to \texttt{Indexed} objects (array accesses) and \texttt{Constant}s. It is important to note that when recursing in an equality, when visiting the LHS we set that sub-tree to be a write (so that if there is an \texttt{Indexed} it is changed accordingly) while leaving the RHS as a read (the default state).

\subsubsection{Indexed translation}

Indexed objects are recreated to use OPS accessor macros instead of regular array accesses. 
The first step is determining the name of the new \texttt{Indexed} object: this is the function name in the case of regular \texttt{Function}s while, if the \texttt{Indexed} object is accessing a \texttt{TimeFunction} the name is composed of the function name and the relevant time accessor (e.g. \texttt{u[t0]} becomes \texttt{ut0}).

We then check if an \texttt{Array} of the same name has already been created in the context and if not, we create it and memorise it to be reused when encountered again. These \texttt{Array} objects are specific to OPS and extend Devito \texttt{Array}s by storing whether they are read-only arrays or not and, if they are, prepend the \texttt{const} modifier to their type definition as it is a requirement when defining the parameters of an OPS kernel.

The indexes used to access the array are then extracted and inserted as a simple text \texttt{Macro} as the index for this new array. This text will contain comma separated integers detailing the shift from the loop value (for example, for \texttt{ut0[x - 1][y + 1]} we would have "-1,1" as the new index). These are not surrounded by the \texttt{OPS\_ACCx} macro call yet as this is indexed depending on the order of the parameters that is not yet known.

Once the expressions are completely translated the parameters are found using the \texttt{FindSymbols} visitor and sorted by type (arrays or scalars) and alphabetical order. This order is then used to insert the correct accessors in the \texttt{Indexed} objects in a similar recursive way to what has been described before. For example, if our kernel signature is \texttt{void Kernel0(const float * ut0, float * ut1)} we will use \texttt{OPS\_ACC0} to access elements of \texttt{ut0} and \texttt{OPS\_ACC1} to access elements of \texttt{ut1}.

\subsubsection{Constants}

\texttt{Constant}s are simply renamed to have the dereference operator at the start of their name. This accomplishes two things: when generating the parameters for the OPS kernel, constants are pointers and when using them in the expressions, they're dereferenced.

\subsection{Header file creation}

Once the expressions have been translated and the parameters identified \texttt{Callable}s are created and inserted in a list within the \texttt{Operator}.
OPS kernels are numbered to avoid name clashes (multiple loops within an operator could potentially be offloaded to OPS).

At compile time the kernels are inserted in a \texttt{List}, an IET node that, as the name suggests, contains a list of nodes. This is then passed to a \texttt{CGen} visitor that generates the code string that is then written to a \texttt{.h} file that is included in the \texttt{.c} file containing the Devito kernel code. Both these files will then be passed to the OPS translator at compile time.

\begin{lstlisting}[language=c, caption=Devito-generated OPS kernel]
void Kernel0(const float * ut0, float * ut1, const float *dt, const float *h_x, const float *h_y)
{
  ut1[OPS_ACC1(0,0)] = -1.0F*(*dt*ut0[OPS_ACC0(0,0)]/((*h_y**h_y)) + *dt*ut0[OPS_ACC0(0,0)]/((*h_x**h_x))) + 5.0e-1F*(*dt*ut0[OPS_ACC0(0,-1)]/((*h_y**h_y)) + *dt*ut0[OPS_ACC0(-1,0)]/((*h_x**h_x)) + *dt*ut0[OPS_ACC0(0,1)]/((*h_y**h_y)) + *dt*ut0[OPS_ACC0(1,0)]/((*h_x**h_x))) + ut0[OPS_ACC0(0,0)];
}
\end{lstlisting}

\section{Invoking OPS kernels}
Once OPS Kernels have been generated, the next step is to invoke these kernels as part of a parallel loop.
In OPS, this means invoking \texttt{ops\_par\_loop} and passing in the kernel to be used, the iteration space, the starting indices in all dimensions and the arguments required by the Kernel.

In order to achieve this, various code generation features were added to Devito. Among these are the ability to nest function calls, the introduction of new data types and the possibility of initialising symbols with the result of function calls. These will be covered in more detail in the next sections.

\subsection{Data declaration}

The first thing to do is transform the data in \texttt{ops\_arg}s that can be passed to OPS kernels.

There are two types of \texttt{ops\_arg}s: datasets and globals. Datasets represent the main data that OPS operates upon (in the context of Devito, these roughly correspond to \texttt{Function}s) while globals can be small constant arrays or scalars.

Globals can be passed directly to kernels, while datsets need to be transformed into \texttt{ops\_dat}s first.

\subsection{ops\_dat creation} \label{opsdat}

\subsubsection{TimeFunction}

When talking about how Devito functions map to OPS datasets it is important to make a distinction between \texttt{Function}s and \texttt{TimeFunction}s. This is due to the fact that the outer time loop is not offloaded to OPS and therefore \texttt{TimeFunction}s at different time steps need to be treated as separate datasets.

Calculations of the value of some time function $u$ at time $t$ rely on the values at previous time steps. In order to access the needed states the options are to either store all the computed values at different time steps or to only store the required states (for example, if the only needed state is the one at the previous time step, we would store $u(t)$ and $u(t-1)$) and rotate through them as needed when computing future states, discarding the ones that are no longer needed. 

If we are storing $u(t)$ and $u(t-1)$, this means that when calculating $u(t+1)$ (using the values at $u(t)$ we can overwrite the data for $u(t-1)$ as this is no longer needed. Practically, this means that to store $u$ in memory an array of shape $(2, size x, size y)$ (assuming a 2D grid) is used and that the time indexes used are rotated at every iteration. The indexes are calculated as $t0 = time \mod n$, $t1 = (time + 1) \mod n$, ..., $t(n-1) = (time + n - 1) \mod n$ where $n$ is the amount of time steps stored and $time$ is the actual time step value. 

Going back to the 2D $u$ example, at every time iteration \texttt{u[t1] = foo(u[t0])} is calculated using these time accessors with the actual values of \texttt{t1} and \texttt{t0} rotating between $0$ and $1$ (and therefore not requiring any other time step besides the ones used in the expression).

As seen previously the translated expression in the OPS kernel would take these as two separate array arguments, \texttt{ut0} and \texttt{ut1} (i.e. they are treated as separate datasets), and therefore multiple \texttt{ops\_dat} need to be created for each time step used in the kernel. 

The creation of the actual \texttt{ops\_dat}s is then done similarly to how it's done for normal \texttt{Function}s with the difference that the information relative to the time dimension is removed.

In order to pass the correct pointer to the \texttt{ops\_decl\_dat} call a new Devito type, \texttt{FunctionTimeAccess} has been created. This represents a pointer to the array representing a \texttt{TimeFunction} at a specific time index (e.g. \texttt{\&u[0]}).

\subsubsection{Extracting relevant information about Functions}

In order to create an \texttt{ops\_dat} the following information is required:

\begin{itemize}
    \item Array size
    \item Base indexes
    \item Padding in the negative direction
    \item Padding in the positive direction
    \item C type
\end{itemize}

All this information can be extracted from data held by the Devito representation of the function, with the only manipulation required being having to change the sign for the padding in the negative direction, since Devito represents this with positive integers while OPS uses negative integers.

Most of this information needs to be passed to OPS as arrays of integers. These can be created by adding \texttt{Expression}s to the IET containing \texttt{Sympy} equations (\texttt{Eq}) with a \texttt{SymbolicArray} (a new type representing a locally initialized array) on the left-hand side and a \texttt{ListInitializer} on the right-hand side.
A \texttt{ListInitializer} is an already existing construct in Devito representing inline list initializations of arrays (e.g. \texttt{\{1, 2, 3\}}) that has been upgraded to allow integers (previously it would only allow strings or \texttt{Sympy} expressions).

Once all the required array are created the next step is to actually create the \texttt{ops\_dat}. This is done using an \texttt{Element}, an IET node that can contain an arbitrary \texttt{cgen}\cite{cgen} element.
In this specific case, we use an \texttt{Initializer}, a \texttt{cgen} node representing an initialization statement of the form \texttt{type name = value;}.

On the left hand side is an \texttt{OPSDat} symbol and, on the right hand side, the \texttt{Call} node representing the call to \texttt{ops\_decl\_dat}. 

In the case of \texttt{TimeFunction}s we will have to create multiple \texttt{ops\_dat}s. To do this, we create a 1D \texttt{SymbolicArray} of size equal to the amount of time indexes used. We then initialize all its elements as previously described.

It is important to note that Devito finds the symbols that need to be passed as parameters to the \texttt{Operator} by visiting the IET. Therefore, since the symbols representing these \texttt{Function}s are no longer part of an \texttt{Expression} containing a \texttt{Sympy} equation functionality to find these symbols from the arguments of calls within an \texttt{Element} has been added.

The resulting code, once generated, will look as follows:

\begin{lstlisting}[language=C]
int u_dim[2] = {3337, 3337};
int u_base[2] = {0, 0};
int u_d_p[2] = {2, 2};
int u_d_m[2] = {-2, -2};
ops_dat u_dat[2];
u_dat[0] = ops_decl_dat(block_0,1,(int *)u_dim,(int *)u_base,
    (int *)u_d_m,(int *)u_d_p,&u[0],"float","ut0");
u_dat[1] = ops_decl_dat(block_0,1,(int *)u_dim,(int *)u_base,
    (int *)u_d_m,(int *)u_d_p,&u[1],"float","ut1");
\end{lstlisting}

Note that in this snippet $u$ is a \texttt{TimeFunction}, therefore we have two \texttt{ops\_dat}s to describe the data at each relevant timestep.

\subsubsection{IET visitor extension: visiting \texttt{Element}s}

After the IET is ready Devito will derive the parameters for the \texttt{Operator} by visiting the IET using a \texttt{FindSymbols} \texttt{Visitor}. This visitor has three modes of operation: \textbf{symbolics} (collects \texttt{AbstractSymbol} objects), \textbf{free symbols} and \textbf{defines} (collects bound objects - variables initialized within the IET).

The parameters will then be the union of the results from running \texttt{FindSybols} in \textbf{symbolics} and \textbf{free symbols} excluding the symbols found in \textbf{defines} mode.

However, since Devito did not have the capability to find these in \texttt{Element}s (due to them being able to contain different \texttt{cgen} nodes and not having any need for this before) when \texttt{Expression}s offloaded to OPS are substituted with OPS API calls any symbol used in these is not found. This results in \texttt{undeclared identifier} errors at compile time, as these symbols are used but never passed to the \texttt{Operator} C function. 

To overcome this \texttt{Element} nodes need to be upgraded to be able to find the symbols contained inside it. This has been done for \texttt{cgen.Initializer}s by simply defining the relevant properties to either return the appropriate values taken from the \texttt{Node} on the RHS or the LHS in the case of \textbf{defines}.

\subsection{Parallel loop invocation}

Once all the datasets are initialized the next step will be to invoke the \texttt{ops\_par\_loop} within the time loop. This replaces the space loop and it's what tells OPS to perform the computation.

\texttt{ops\_par\_loop} takes as arguments the OPS kernel, the number of dimensions, the iteration space, and $N$ \texttt{ops\_arg}s representing the arguments to the kernel. These will be created from the \texttt{ops\_dat}s initialized outside the loop and from the already known constants.

\subsubsection{Kernel pointer}

The first argument to \texttt{ops\_par\_loop} is a function pointer to the OPS kernel function to be used in the loop. To pass this in a \texttt{FunctionPointer} type has been created that simply prints out the function name.

\subsubsection{Iteration space}

The iteration space can be extracted from the \texttt{IterationTree} containing the expressions being moved in an OPS kernel. These are then used to initialize a \texttt{SymbolicArray} in a similar way to what is described in \ref{opsdat}, with lower and upper bounds for each dimension being the elements of the array.

\subsubsection{Kernel arguments}

Kernel arguments are passed to the \texttt{ops\_par\_loop} call by nesting in it calls to \texttt{ops\_arg\_dat} in the case of datasets and \texttt{ops\_arg\_gbl} in the case of small global scalars or arrays.

In the case of globals, we pass to \texttt{ops\_arg\_gbl} a pointer to our data, the size (in our case, since the only globals we have are scalars, we are going to default to 1), a string representing the C type of the data and whether the data is read and/or write (again, we are dealing with constant scalars, therefore we will default to \texttt{OPS\_READ}).

For datasets, we will have to pass the relevant \texttt{ops\_dat}, an \texttt{ops\_stencil} (defining what neighbouring data points are used during execution of the kernel over a point) and, similarly to the globals case, the C type (in string format) and the R/W setting.

In the case of \texttt{TimeFunction}s arguments, as seen before, we have multiple \texttt{ops\_dat}s in an array. To index the array the time indexing variables are used so that we rotate through the datasets correctly at every time step.

\subsubsection{Stencil creation}

OPS stencils define what data points are accessed in a dataset when the kernel is run for a single data point. 

Taking a Devito expression before OPS translation as an example, we have that for
\begin{equation*}
    \texttt{u[t1][x][y] = u[t0][x][y] + u[t0][x - 1][y]}
\end{equation*}

the OPS stencils would contain the following offsets: 
\begin{equation*}
\begin{aligned}
    stencil_{u[t1]} &= \{ (0, 0) \} \\
    stencil_{u[t0]} &= \{ (0, 0), (-1, 0) \}    
\end{aligned}
\end{equation*}

To create these stencils we loop over all the \texttt{Indexed} objects in an expression and analyse their accessors. To find them \texttt{retrieve\_indexed} is used, a function already present in Devito that simply visits the equation graphs and returns a collection containing all the found \texttt{Indexed}.

For each function that is indexed a \texttt{set} is created. This will contain tuples of tuples containing the dimension being accessed and its offset.
Looking at our previous example, we would have \texttt{accesses[u] = \{((t0, 0), (x, 0), (y, 0)), ((t1, 1), (x, 0), (y, 0)), ((t1, 0), (x, -1), (y, 0))\}}.

This is done for every equation that will go in the kernel so that for every function we will have a complete set of all accesses across equations.
The next step is then to generate \texttt{ops\_stencil} objects. This is done through a call to \texttt{ops\_decl\_stencil} that takes the dimensions, the number of points, an array containing a flattened list of these points and the stencil name as a string. The first two arguments are easily derived from the accesses set, while the name is selected using a simple convention:  \textbf{s\{dimensions\}d\_\{function\_name\}\_\{points\}pt}.

The array is created by iterating over the set of points for a function and extracting the offset values from the space dimensions and appending them at the end of a list that is then passed to a \texttt{ListInitializer} used to initialize an \texttt{SymbolicArray} of integers in a Devito \texttt{Expression}.
The \texttt{OPSStencil} symbol is then initialized with a \texttt{cgen.Initializer} with the \texttt{Call} to \texttt{ops\_decl\_stencil} on the RHS as previously described and then stored for later use in the \texttt{ops\_arg\_dat} call.

In the case of \texttt{TimeFunction} we first need to split grouping the element by the accessor used in the time dimension and then proceed as previously detailed.
Looking at the earlier example, the sets to be used would be \texttt{ut0 = \{((x, 0), (y, 0))\}} and \texttt{ut1 = \{((x, 0), (y, 0)), ((x, -1), (y, 0))\}}.

For a diffusion operator with space order 2 the following stencils will be generated:

\begin{lstlisting}[language=c]
int s2d_ut0_5pt[10] = {0, 1, 1, 0, 0, -1, -1, 0, 0, 0};
ops_stencil S2D_UT0_5PT =
    ops_decl_stencil(2,5,(int *)s2d_ut0_5pt,"S2D_UT0_5PT");
int s2d_ut1_1pt[2] = {0, 0};
ops_stencil S2D_UT1_1PT =
    ops_decl_stencil(2,1,(int *)s2d_ut1_1pt,"S2D_UT1_1PT");
\end{lstlisting}

\subsubsection{IET visitor extension: nested calls}

Devito doesn't natively support nested calls (e.g. \texttt{foo(bar())}) therefore functionality to do so has been added during the course of this project.

The first step is enable the nested calls to be visited by IET visitors: this is done by defining the \texttt{children} property of the \texttt{Call} class to return a list of all arguments that are instances of \texttt{Call}.

Then, a \texttt{visit\_Call} needs to be created or modified for relevant visitors: these define the logic when a \texttt{Call} node is visited.

The \texttt{CGen} visitor, used to generate the actual C code has been updated so that when the argument string for the call is generated in the \texttt{\_args\_check} method we recurse down when a \texttt{Call} argument is encountered and the resulting code string for that subtree is appended to the argument list.

Similarly, the \texttt{FindSymbols} \texttt{visit\_Call} method has been updated so that instead of simply applying the search rule to the current \texttt{Node}, the \texttt{Call} arguments are visited and the results found are added to the set of symbols.

By default, Devito assumes that a function call is not nested and that its result is not used, therefore it always uses a \texttt{cgen.Statement} (that adds a semicolon at the end of the generated text). However, when nesting calls we do not want the \texttt{Call} text to be rendered as a C statement, but as simple text: to define when this is the case a \texttt{semicolon} field has been added to the \texttt{Call} class to be used in the \texttt{CGen} visitor.
This is also useful when initializing data using a \texttt{cgen.Initializer} to avoid double semicolons on the initialization line.

\section{Summary}

This chapter has shown how to generate:
\begin{itemize}
    \item OPS kernels living in a header file that use OPS accessor macros
    \item Declarations of OPS datasets, stencils and iteration ranges
    \item Invocations of OPS kernels with the appropriate arguments
\end{itemize}

This is what is required to use the OPS source-to-source translator, compile the resulting code and link it with the OPS libraries.

\chapter{Compilation and execution}

This chapter describes how OPS code is compiled and executed by Devito.

First the current compilation and execution pipelines for both Devito and OPS are described in order to understand their differences, then a description on how the two have been integrated is presented.

\section{Devito compilation and execution pipeline}
\label{sec:devitocomp}
When an user tries to execute an \texttt{Operator} by invoking the \texttt{apply} method the compilation process will be triggered (if it's the first execution of the operator). This will involve compiling the Devito generated code into a shared library, loading it into Python and executing the entry point as defined by Devito.

\subsubsection{Code generation}

In order to compile our code the first step is transforming the IET into a C code string. This is done by retrieving the \texttt{ccode} property of the \texttt{Operator}. This property uses the \texttt{CGen} visitor to generate C code from an IET node such as \texttt{Operator}. This will result in a string containing a complete and valid C file containing headers, type declarations and a function representing the entry point of the computation.

\subsubsection{Compilation and execution}

This C string is then passed to the \texttt{jit\_compile} method that performs just-in-time (JIT) compilation by calling the \texttt{compile\_from\_string} function from the \texttt{codepy} \cite{codepy} library. In order to compile a \texttt{Compiler} object (that extends \texttt{codepy}'s \texttt{GCCToolchain} class) is passed to \texttt{compile\_from\_string} to define what compiler to use and what compilation flags are needed. This will generate a \texttt{.so} that can then be loaded in Python using Numpy's C-Types Foreign Function Interface \cite{numpyctypeslib}. Once loaded, the function can be simply executed as a normal Python function.

\section{OPS compilation pipeline}

OPS programs have a two step compilation process. The first step is running the OPS translator, a source to source compiler that, given some \texttt{C++} source code that uses the OPS API, generates code that can target multiple platforms including CUDA, MPI, OpenMP, OpenACC and others. 
The second step compiles the generated code and links it to the OPS libraries to create executables for the different platforms targeted.

The standard OPS workflow to execute these two steps is by using \texttt{Makefile}s. For any new program, an user would simply create its own \texttt{Makefile} that includes OPS provides makefiles and sets a few variables describing where the code lives as shown in the example Makefile in listing \ref{lst:makefileops}. Running \texttt{make} will then start the compilation process and generate all the executables needed.

The assumptions that both the translator and the Makefiles make is that executables for all target platforms need to be generated. This renders the Makefiles quite complex to unravel if we only want to compile for a single platform manually (or, in our case, automatically within Devito) but, from a user perspective, are quite straight forward to use.

\begin{lstlisting}[float, language=make, caption={An example Makefile for an OPS application},label={lst:makefileops}]
include $(OPS_INSTALL_PATH)/../makefiles/Makefile.common
include $(OPS_INSTALL_PATH)/../makefiles/Makefile.mpi
include $(OPS_INSTALL_PATH)/../makefiles/Makefile.cuda
include $(OPS_INSTALL_PATH)/../makefiles/Makefile.hdf5

HEADERS=diffusion_so$(SPACE_ORDER).h

OPS_FILES=diffusion_so$(SPACE_ORDER).c

OPS_GENERATED=diffusion_so$(SPACE_ORDER)_ops.cpp

APP=diffusion_so$(SPACE_ORDER)
MAIN_SRC=diffusion_so$(SPACE_ORDER)

include $(OPS_INSTALL_PATH)/../makefiles/Makefile.c_ap
\end{lstlisting}

\section{OPS compilation in Devito}
\label{sec:integration}

An integration that allows to run OPS code directly from Devito has been successfully implemented as part of this project. However, for the CUDA case, it relies a minor change to how the OPS libraries are built.

\subsubsection{Changes to OPS}

The OPS build process has been altered to allow the OPS CUDA libraries to allow their use in the creation of shared libraries that can be loaded in Python. The required change was simply to add the \texttt{-fPIC} flag to the \textbf{nvcc Xcompiler} list of flags used to build the OPS libraries.

The \texttt{fPIC} flag tells the compiler to generate position-independent code, so that the generated machine code does not depends on being at certain locations (for example, for jumps) and therefore it can be used for shared libraries and dynamic linking.

\subsubsection{Compiling in Devito}
\label{sec:compilingindevito}
When compiling in Devito the generated OPS code is simply written to files and, using the \texttt{subprocess} Python module the OPS translator is invoked on them to generate all the required source files for different target platforms.

Then, depending on the \texttt{DEVITO\_OPS\_TARGET} environment variable that defines what platform to target, the generated files are used to create a shared library that can be loaded and executed by Devito.

For most target platforms the process is then very similar to what is described in section \ref{sec:devitocomp}, with the difference that we are compiling multiple files and the code strings passed to the \texttt{compile\_from\_string} have to be read from disk, since what we are now compiling are the OPS translator generated files and not the Devito generated code. For each OPS target platform a new \texttt{Compiler} class has been created that replicates what the OPS Makefile does for that platform.

\subsubsection{CUDA compilation}

Compiling for CUDA differs slightly in the fact that device code (the actual CUDA kernel) needs to be compiled with \texttt{nvcc} first and the resulting object linked when compiling the host code. The compilation pipeline therefore looks a bit different in this case.

Two \texttt{Compiler} classes have been created, one for the host (\texttt{OPSCUDAHostCompiler}) and one for the device (\texttt{OPSCUDADeviceCompiler}). These two compilers are then used to generate object files for the CUDA kernel and the C++ code respectively by setting the \texttt{objcet} parameter to \texttt{True} when calling \texttt{compile\_from\_string}. The two objects are then linked using the \texttt{link\_extension} method on the host compiler, passing in the two \texttt{.o} files.

\begin{figure}
    \centering
    \includegraphics{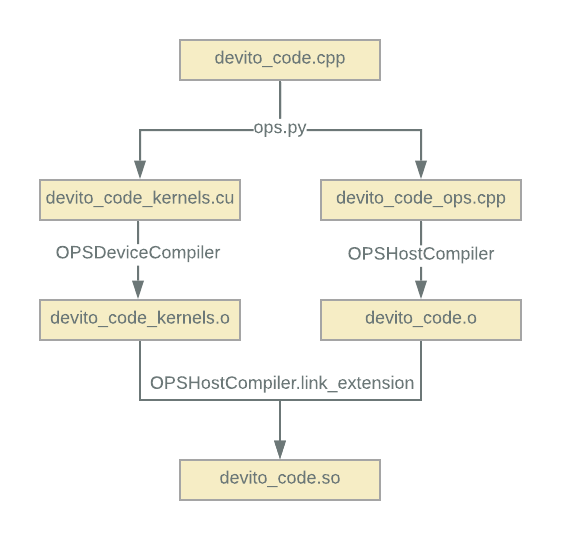}
    \caption{CUDA compilation pipeline in Devito}
    \label{fig:cudacomp}
\end{figure}

\subsubsection{OPS translator concerns}
\label{sec:translatorconcerns}
The biggest obstacle to a clean integration is the OPS translator. Currently this component has two big flaws: it uses Python 2.7 (that will stop being maintained in 2020 \cite{python2}) and it is organized as a monolithic script for standalone execution, making it cumbersome to use within Devito.

Upgrading the translator to Python 3 would be a good first step in simplifying its integration within Devito, but a better solution would be a complete rework of the OPS translator as a proper Python module. It would need to expose methods to translate OPS code to user defined target platforms without having to touch the file system (currently, it generates code for all platforms supported by OPS and writes all the resulting files to disk) to grant the user more refined control.

This would allow easier use of the translator within Devito to target specific platforms using OPS, facilititating the use of Devito \texttt{Compiler}s that currently requires Devito to read the generated files from disk.

%% file: evaluation/evaluation.tex
\chapter{Evaluation}
\label{cha:eval}

When deciding whether to adopt a new library into an existing established software project there are multiple factors to consider. These vary depending on the application and library examined.

In the case of Devito and OPS the question that need to be answered are
\begin{itemize}
    \item Is OPS a solid enough piece of software to be integrated in the Devito environment?
    \item How hard it is (and therefore what is the development cost) to integrate OPS inside Devito?
    \item Does the performance provided by OPS (primarily on GPU) justify the develoment cost of integrating the library into Devito?
\end{itemize}

Answering these questions is the ultimate goal of this work, in order to help Devito developers decide on the best path for the evolution of Devito and the adoption of new architectures such as GPUs. 
Possible ways forward include further integration of OPS or similar tools or generating GPU code in house within Devito without having to rely on external libraries: this would give Devito the benefit of tailoring the generated code to the specific class of problem it targets but add an extra cost in terms of maintaining this new code generation pipeline and implementing optimizations directly.

In this chapter we evaluate the performance of Devito + OPS on GPU architectures using CUDA. This will allow the reader to have an indication of what performance could be expected by using OPS as the GPU backend for Devito and will be a key element in informing a decision on whether OPS should be adopted or not.
\section{Performance evaluation}

\subsection{Hardware used}

\subsubsection{CUDA benchmarks}

To run CUDA benchmarks, a dedicated machine has been used. The specifications are as follows:

\textbf{CPU}: Intel Core i7-4770k\\
\textbf{GPU}: NVIDIA GTX 1080\\
\textbf{CUDA driver version}: 9.1

Specifically, the GPU specifications are as follows \cite{nvidia-gtx-1080-specs}:\\
\textbf{CUDA Cores}: 2560\\
\textbf{Graphics clock}: 1607 MHz\\
\textbf{Processor clock}: 1733 MHz\\
\textbf{Memory size}: 8 GB\\
\textbf{Memory interface width}: 256 bit\\
\textbf{Memory bandwith}: 320 GB/s\\
\textbf{Single Precision performance (theoretical)}: 8228 GFLOPS/s

\subsubsection{OpenMP benchmarks}

To run OpenMP benchmarks an Azure H8 instance has been used. The relevant specifications for this machine are as follow \cite{azurehpcsizes}:

\textbf{CPU}: Intel Xeon E5 2667 v3 (8 vCPU)\\
\textbf{Memory}: 56 GB\\
\textbf{Memory bandwith}: 40 GB/s

\subsection{Examined problem}

The problem we used to benchmark the OPS backend is the diffusion equation, a simple differential equation that makes use of all the implemented features in the Devito OPS backend and is complex enough to give a good indication of what the performance would be with more common and complex seismic imaging problems.

The diffusion equation is expressed as follows:
\begin{equation*}
    \frac{\partial u}{\partial t} = \nu \frac{\partial ^2 u}{\partial x^2} + \nu \frac{\partial ^2 u}{\partial y^2}
\end{equation*}

and can be expressed in Devito as such:
\begin{lstlisting}
grid = Grid(shape=(nx, ny))
u = TimeFunction(name='u', grid=grid, time_order=1, space_order=so)
eqn = Eq(u.dt, v * (u.dx2 + u.dy2))
stencil = solve(eqn, u.forward)
op = Operator(Eq(u.forward, stencil))
\end{lstlisting}

For the purposes of our benchmarks, the main parameter will be the space order, as changing it leads to generating more operational intensive and complex code to have a range of useful results. 

Other parameters that will be varying are the grid size and the Devito Symbolic Engine (DSE) optimization level. All benchmarks will run for 1000 timesteps.

\newpage
\subsection{Benchmarking methodology}

\textbf{opescibench} \cite{opescibench} has been used to run our benchmarks and plot our results. It is a tool specifically designed to benchmark Devito operators, providing objects to run parametrized operators, collect performance results and plot them.

In order to enable profiling of informations such as operational intensity and GFLOPS/s advanced profiling has been enabled by setting the \texttt{DEVITO\_PROFILING} environment variable, as the default profiling setting only collects timing information.

To benchmark the Devito core backend the latest official version of Devito (at the time of writing) has been used. This can be found on GitHub at \url{https://github.com/opesci/devito}. The OPS backend has been developed independently on a fork of the Devito repository, as the explorative nature and time frame of the project did not allow it to be merged in the master Devito repository. The Devito+OPS backend repository can be found at \url{https://github.com/vincepandolfo/devito}.

\subsection{Roofline model}

To present our results the roofline performance model \cite{roofline} has been used. It provides an easy to understand visual reference for memory and compute bounds on the observed machine in relation to the operational intensity (FLOP/Byte ratio) of a computation. 

Operational Intensity and GFLOPs for every run are calculated automatically by Devito and are then plotted on the roofline and annotated with the total runtime and the space order (SO) of the run.

For the memory bounds the vendor provided memory bandwidth are used. For the compute bounds, theoretical maximum performance (marked on the plots as \textit{ideal peak}) and measured performance have been used.

To obtain measured performance bounds the \textit{LINPACK} benchmark \cite{linpack} and a \textit{gravitational n-body simulation} \cite{nbody} (bundled with CUDA samples) have been used on the Xeon CPU and GTX 1080 CPU respectively. 

\subsection{CUDA results}

\begin{figure}[h]
\begin{subfigure}{0.5\textwidth}
    \centering
    \includegraphics[width=\linewidth]{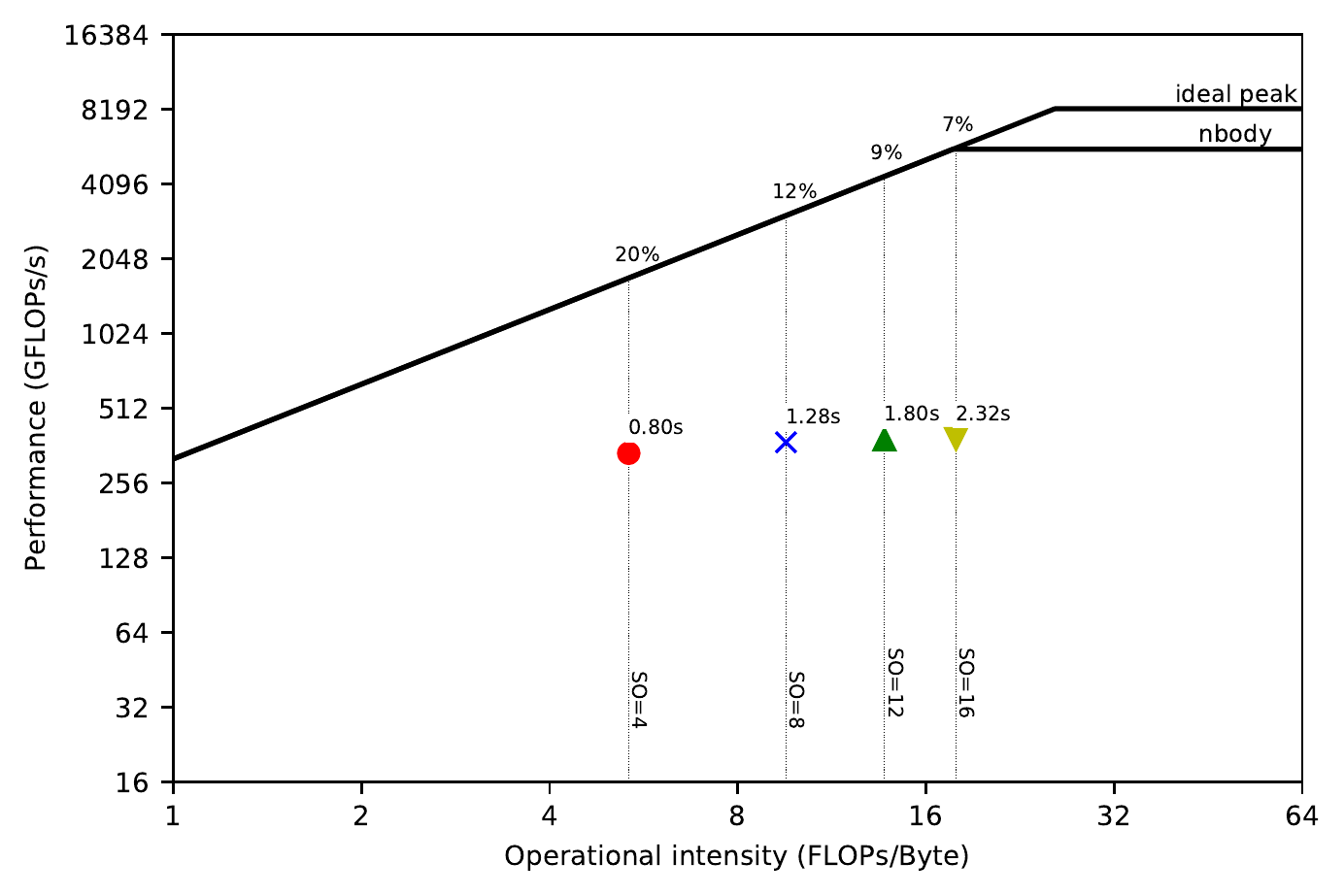}
    \caption{$2500^2$ grid}
\end{subfigure}
\begin{subfigure}{0.5\textwidth}
    \centering
    \includegraphics[width=\linewidth]{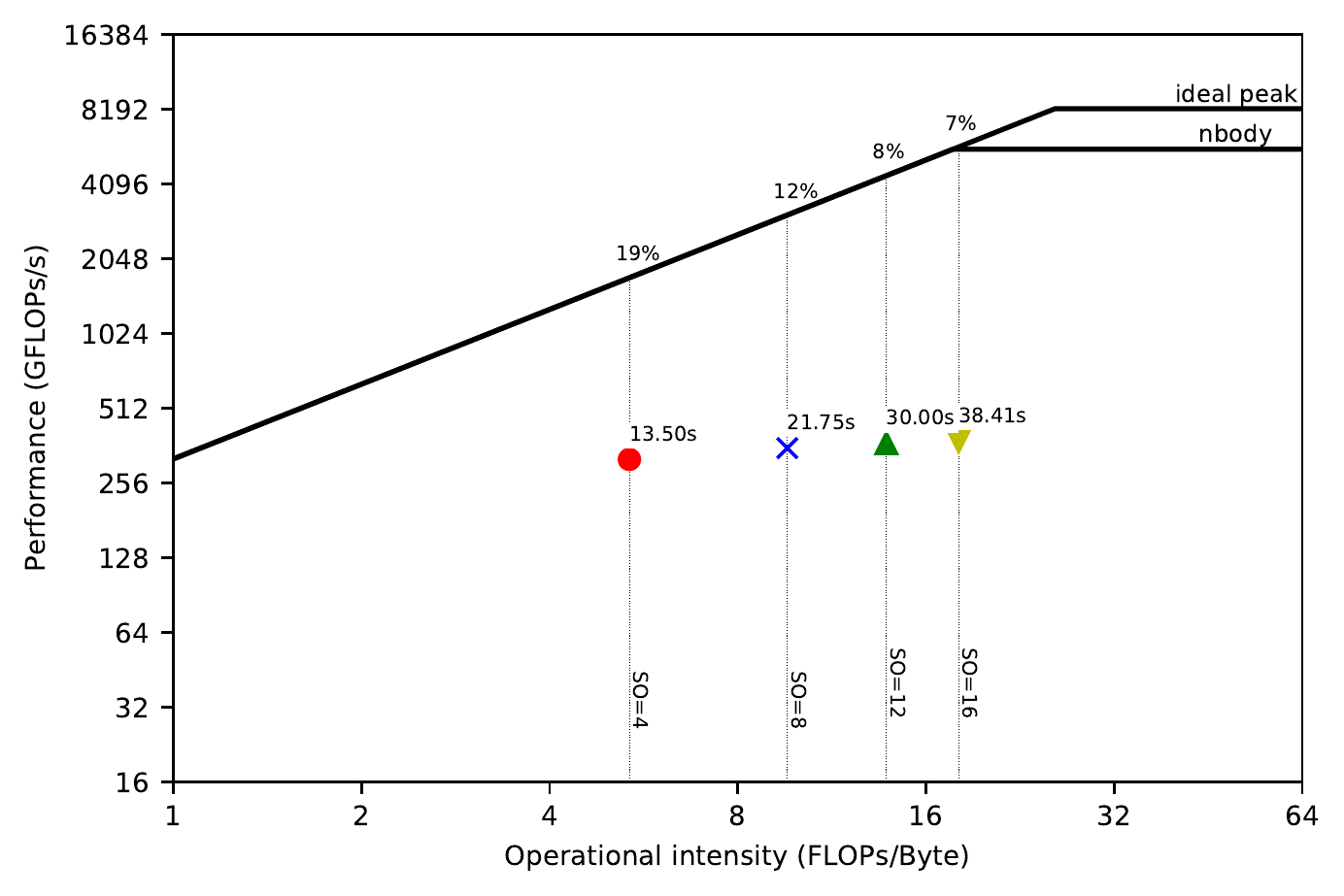}
    \caption{$10000^2$ grid}
\end{subfigure}
    \caption{OPS backend, CUDA, advanced DSE}
    \label{fig:diffcudaadvanced}
\end{figure}

At first, benchmarks were ran only using Devito's default DSE mode, \texttt{advanced}. This produced the results shown in figure \ref{fig:diffcudaadvanced}, with maximum peak utilization of \textbf{20\%} in the $2500\times2500$ grid case and minimum of \textbf{7\%} for the lowest and highest space order respectively. The results for the $10000\times10000$ grid are similar.

After discussion with Devito and OPS developers regarding the possible causes of this it was suggested to optimize the kernels by reducing the number of divisions as it is an extremely expensive operation to perform on the GPU. 

This reduction can be performed by the Devito Symbolic Engine (DSE) in \textbf{aggressive} mode: among other optimizations, it substitutes common divisions with multiplications as shown in listing \ref{lst:so2kerneldivision}. However, for more complex kernels the DSE in aggressive mode would apply further transformations that would not be necessarily beneficial to performance on a GPU.

This is not done by default in Devito as it is not necessary on CPUs, but on GPUs it makes a great difference. This suggests that some exploration on what optimizations are most beneficial is necessary in order to create a GPU specific DSE mode.

\begin{figure}
\begin{subfigure}{0.5\textwidth}
    \centering
    \includegraphics[width=\linewidth]{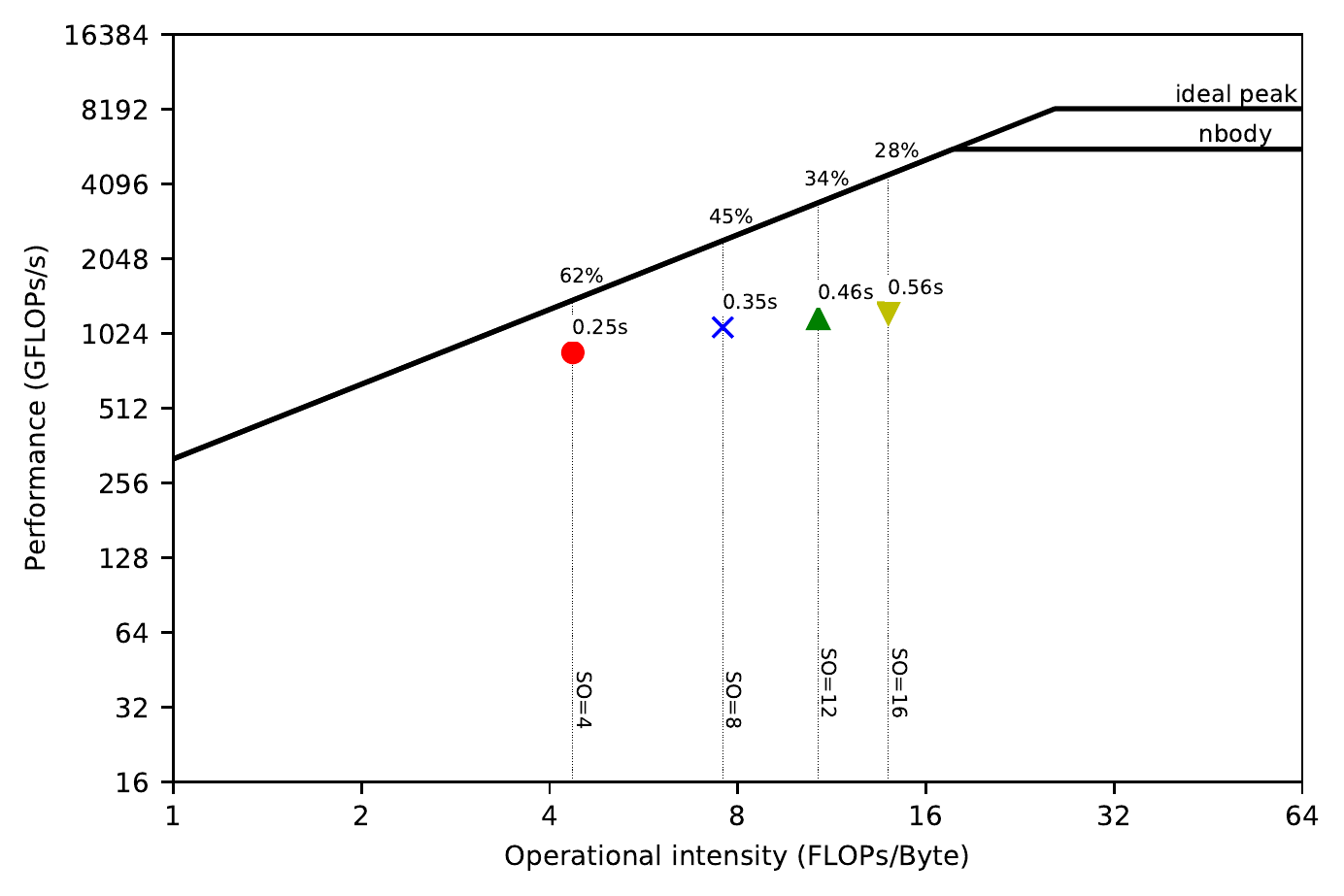}
    \caption{$2500^2$ grid}
\end{subfigure}
\begin{subfigure}{0.5\textwidth}
    \centering
    \includegraphics[width=\linewidth]{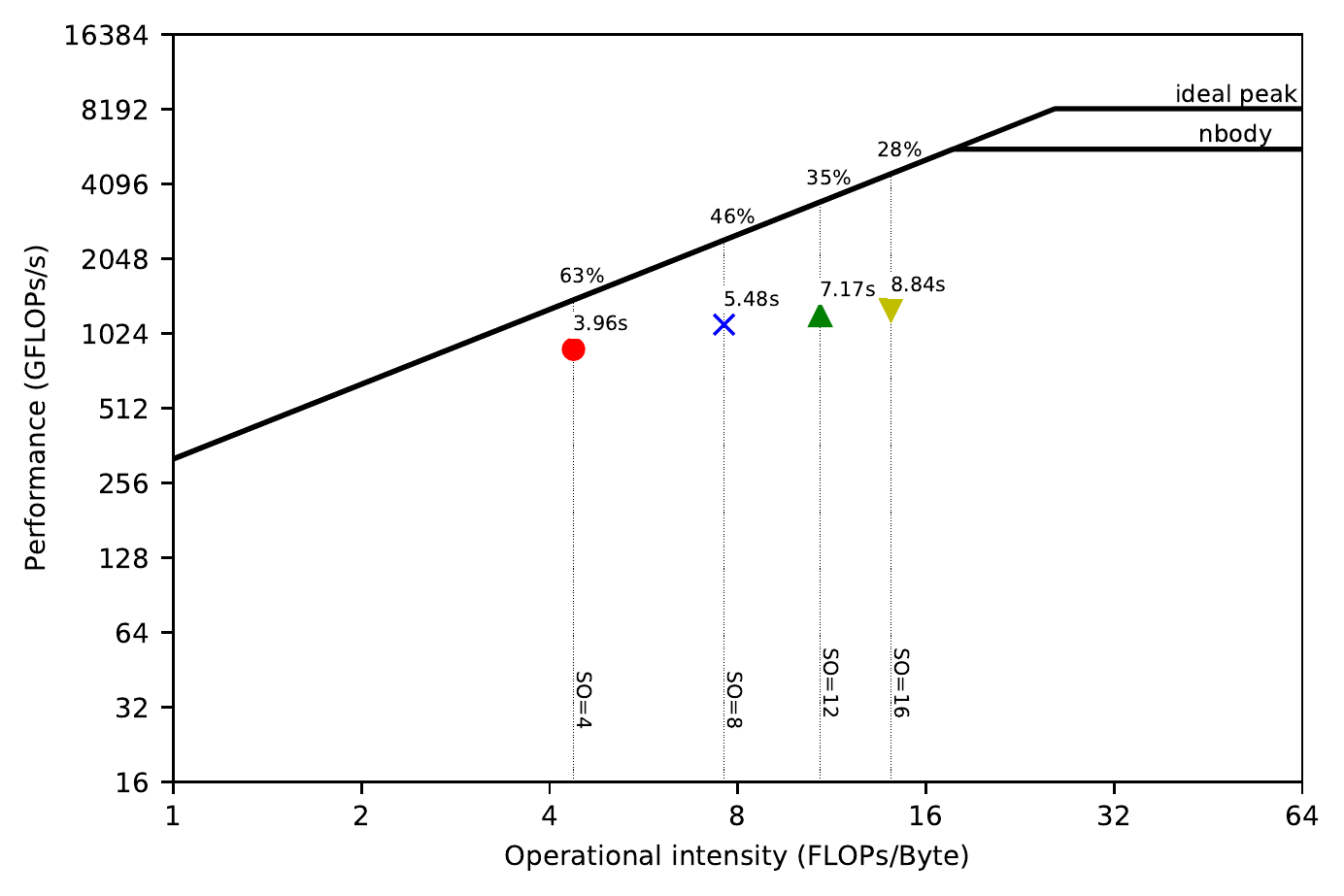}
    \caption{$10000^2$ grid}
\end{subfigure}
    \caption{OPS backend, CUDA, aggressive DSE}
    \label{fig:diffcudaaggressive}
\end{figure}

Running our experiments again with the outlined changes to the kernels we have much better results as shown in figure \ref{fig:diffcudaaggressive}. We now have maximum peak utilization of $62\%$ and minimum of $28\%$ for the $2500\times2500$ case, with similar results on the other grid size. This is between 3 and 4 times more utilization compared to the \texttt{advanced} case, which is a very significant increase in performance.

These results are promising and could warrant further future exploration into OPS by the Devito team.
It is worth noting that the problem examined is simple in nature and does not represent the complexity found in most seismic imaging problems, however this results are still indicative of what performance can be expected in the main loop of these problems. 
\\
\\
\begin{lstlisting}[language=c, label={lst:so2kerneldivision}, caption=2D diffusion kernel with space order 2 (aggressive DSE)]
void Kernel0(const float * ut0, float * ut1, const float *dt
    const float *h_x, const float *h_y)
{
  float r0 = 1.0F / (*h_y**h_y);
  float r1 = 1.0F / (*h_x**h_x);
  ut1[OPS_ACC1(0,0)] = -1.0F*(*dt*ut0[OPS_ACC0(0,0)]*r1 + 
    *dt*ut0[OPS_ACC0(0,0)]*r0) + 5.0e-1F*(*dt*ut0[OPS_ACC0(-1,0)]*r1 +
    *dt*ut0[OPS_ACC0(1,0)]*r1 + *dt*ut0[OPS_ACC0(0,-1)]*r0 +
    *dt*ut0[OPS_ACC0(0,1)]*r0) + ut0[OPS_ACC0(0,0)];
}
\end{lstlisting}

\subsection{OpenMP results}

Since OPS targets multiple platforms, it is worth investigating the performance on CPUs compared to what is already offered by Devito. This is because, if good results are obtained, OPS could potentially substitute the core Devito backend. This benchmark will focus on performance using OpenMP. 

OpenMP is targeted by both the Devito Core and OPS backends, therefore it is easy to draw a performance comparison between the two utilising the same code. In order to enable OpenMP in the Core backend the \texttt{DEVITO\_OPENMP} environment variable must be set to $1$.

Benchmarks have only been run for the $10000 \times 10000$ grid size as a smaller grid size would be completely contained within the cache, therefore not making use of any memory access optimization that might be performed by the different backends.

\begin{figure}[H]
\begin{subfigure}{0.5\textwidth}
    \centering
    \includegraphics[width=\linewidth]{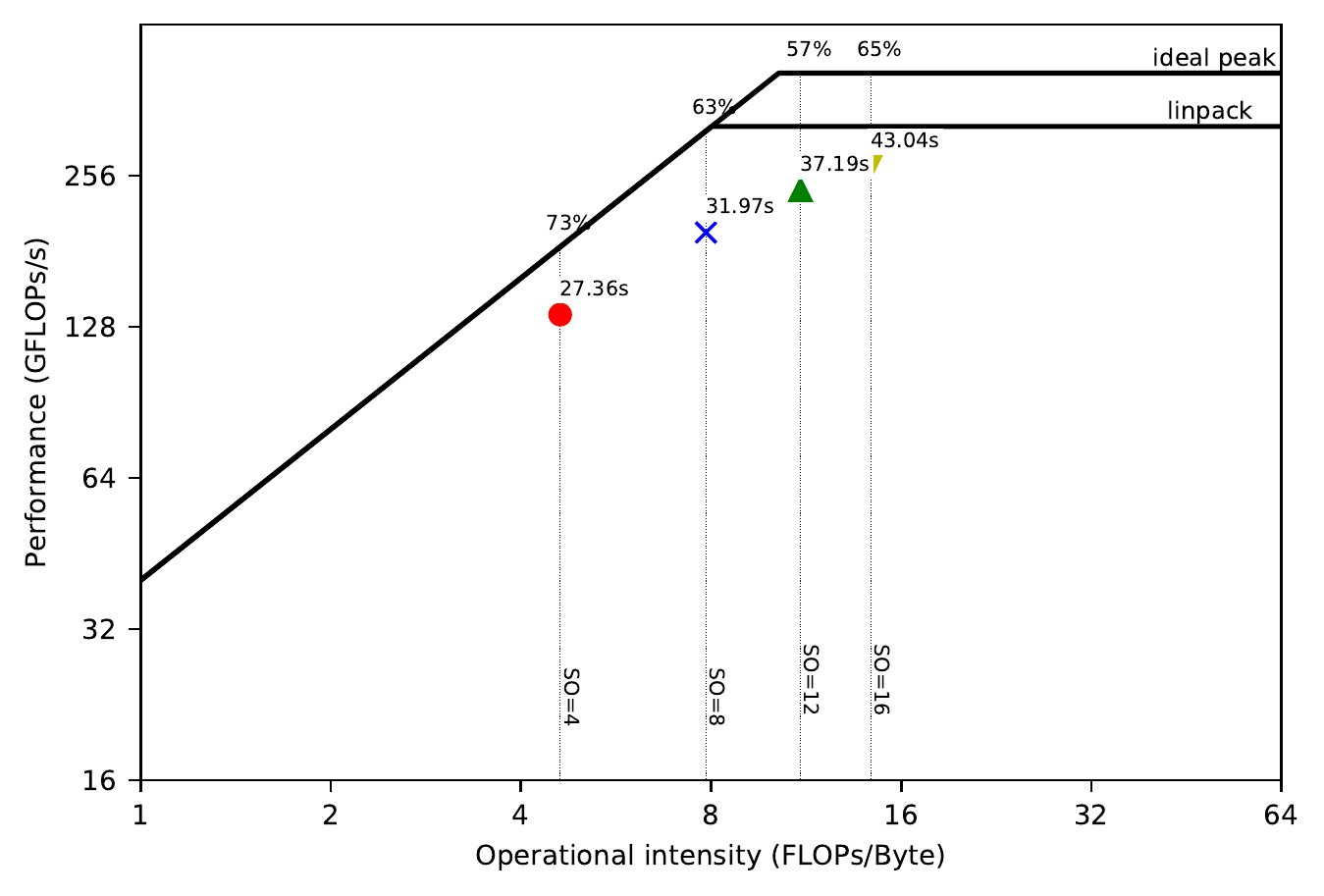}
    \caption{Advanced DSE}
\end{subfigure}
\begin{subfigure}{0.5\textwidth}
    \centering
    \includegraphics[width=\linewidth]{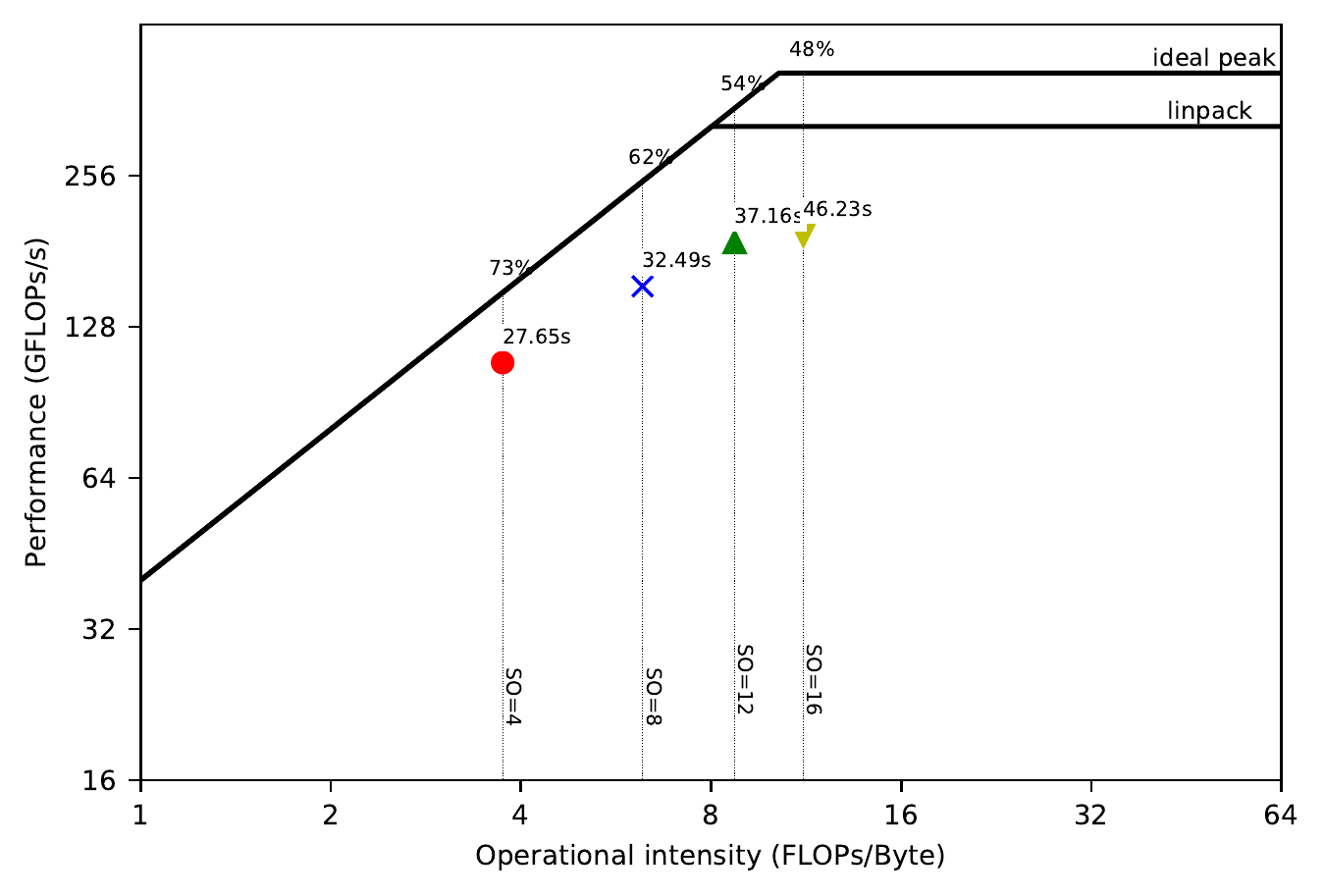}
    \caption{Aggressive DSE}
\end{subfigure}
    \caption{$10000^2$ grid, core backend, OpenMP}
    \label{fig:diffopenmpcore}
\end{figure}

Figure \ref{fig:diffopenmpcore} shows the results for this run, with maximum and minimum peak percentage of $75\%$ (space order 4) and $57\%$ (space order 12) when setting the DSE in advanced mode. As suggested in the previous section, using the aggressive DSE did not provide any benefit in the CPU case, while actually getting worse performance for the highest space order ($48\%$ as opposed to $65\%$).

The same benchmarks have then been run using the OPS backend targeting OpenMP (this is done by setting the \texttt{DEVITO\_OPS\_TARGET} environment variable to \texttt{OpenMP}, as described in \ref{sec:compilingindevito}).

The results were quite disappointing, with minimum and maximum peak performance percentage of $3\%$ and $6\%$ across both DSE modes and runtimes from 10 to 25 times slower.

These results, if confirmed by further exploration and not found to be due to problems in the Devito OPS backend, suggest that it might not be worth substituting the core OpenMP backend with OPS as no benefit would be gained by doing so.

% \textbf{TODO:} work out if these results are actually legit - they're mad low

\begin{figure}[h]
\begin{subfigure}{0.5\textwidth}
    \centering
    \includegraphics[width=\linewidth]{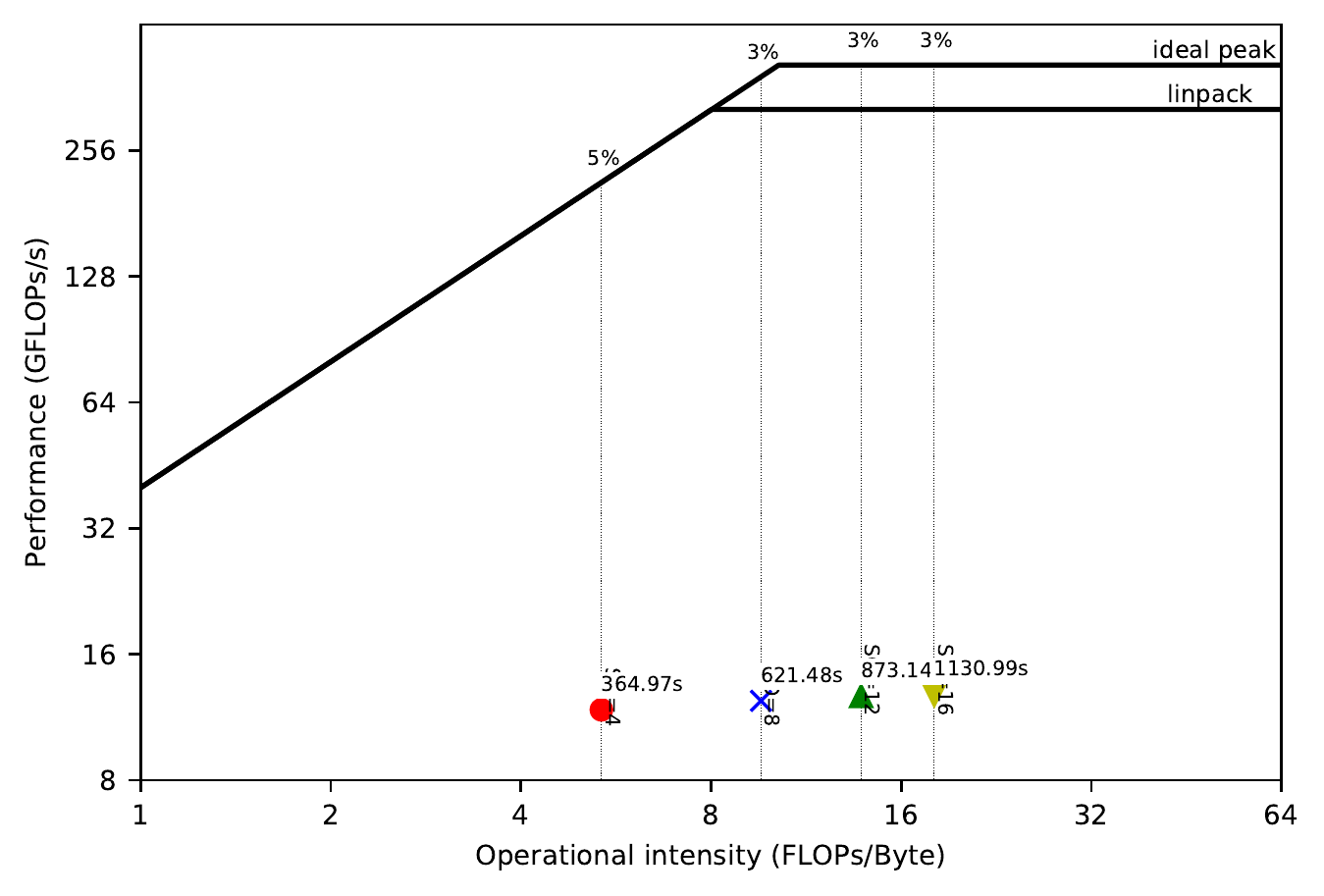}
    \caption{Advanced DSE}
\end{subfigure}
\begin{subfigure}{0.5\textwidth}
    \centering
    \includegraphics[width=\linewidth]{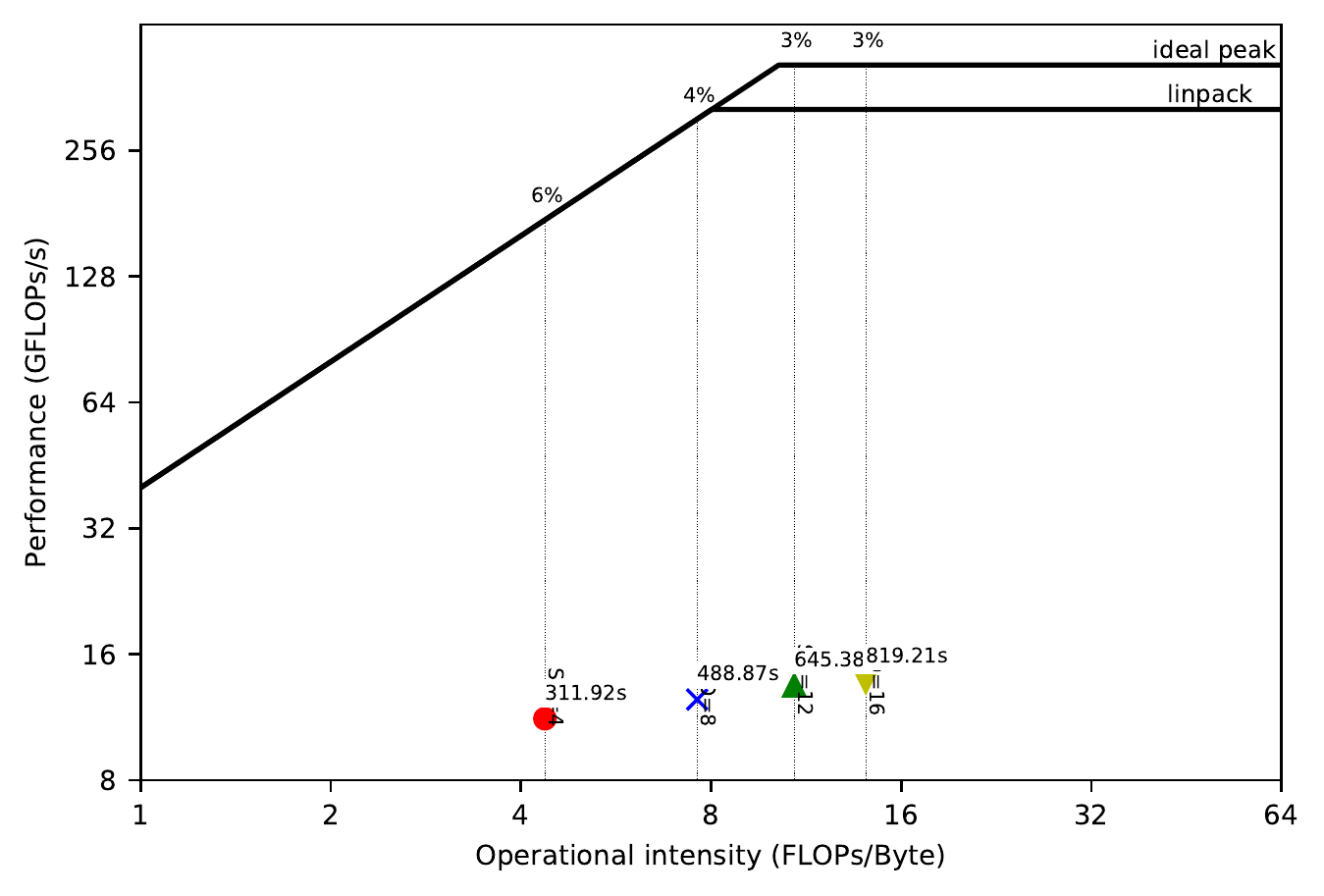}
    \caption{Aggressive DSE}
\end{subfigure}
    \caption{$10000^2$ grid, OPS backend, OpenMP}
    \label{fig:diffcudaaggressive}
\end{figure}

\subsection{Summary}

While the observed performance of the OPS backend on CPU is less than promising, very good results were obtained with CUDA, with good percentages of peak performance and from 4 to 7 times faster execution times compared to the core backend on the hardware used.

\section{Software evaluation}

One of the considerations one has to make when adopting a new library is its soundness as a piece of software. In the case of this project this is even more important, as the investigated library would constitute a critical part of the GPU backend if adopted by Devito.

The questions we will try to answer in this section are:
\begin{itemize}
    \item Is OPS being actively developed?
    \item Is it easy to use?
    \item Is it well engineered?
    \item Does it deliver the required features?
\end{itemize}

Answering these questions will give an important insight on the value of OPS not only as a performance library, but as a usable piece of software.

\subsubsection{Is OPS being actively developed?}
\begin{figure}[h]
    \centering
    \includegraphics[width=\linewidth]{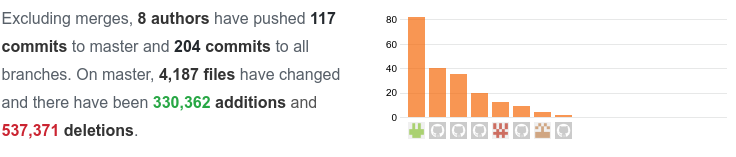}
    \caption{GitHub insights for the OPS repository (accessed 15th Jun 2019)}
    \label{fig:githubinsights}
\end{figure}

\begin{addmargin}{1em}
Yes, as evidenced by the GitHub insights in figure \ref{fig:githubinsights}.
\end{addmargin}

\subsubsection{Is it easy to use?}

\begin{addmargin}{1em}
As mentioned in section \ref{sec:translatorconcerns}, there are some concerns to the usability of the OPS translator: the main one being the fact that it is in Python 2.7 and therefore not directly usable by Devito (which is written in Python 3). Other usability concerns are the fact that it acts as a black box with no control given to the user who is therefore forced to manually read and manipulate files instead of being able to select what code they need to have generated and get it in a convenient form (such as a string).

These problems could be caused from the fact that OPS has probably been developed as a DSL for complete applications and has not been planned to be used within other code generation libraries. 

However, it would not be too difficult for an experienced Python developer to create new interfaces that can provide this kind of functionality and therefore increase usability of the OPS translator within other code generation libraries such as Devito.
\end{addmargin}

\subsubsection{Is it well engineered?}

\begin{addmargin}{1em}
The internal structure of the translator reflects its nature as black box as it is structured as a script as opposed to a library, this is possibly due to the nature of OPS as a DSL for complete applications as opposed to an intermediate representation for other libraries. It is separated in multiple files, each containing a function that generates source code for a target platform. These functions are hundreds of lines long and mostly reflect the final content of the generated files (minus the parts that are changed depending on the user input).

The code is therefore quite simple to understand as it's a sequence of string concatenations, but the sheer size of it makes it difficult to maintain. This design could be improved by introducing OOP concepts to create a higher level definition (such as an abstract syntax tree) of what the generated code looks like. This could be more easily achieved by exploiting a code generation library such as \texttt{cgen} \cite{cgen}.
\end{addmargin}

\subsubsection{Does it deliver the required features?}

\begin{addmargin}{1em}
As shown in the previous section, while the results with OpenMP weren't promising, OPS has shown good performance results on GPUs using CUDA which is what we are most interested in. Therefore we would say that it does deliver the features required by Devito.
\end{addmargin}

While the software quality of some components of OPS might not be excellent, it is definitely possible to improve it significantly without excessive effort. Even at the current state, as shown in section \ref{sec:integration}, integration within Devito is not excessively complicated. OPS does provide the features needed by Devito to target GPUs and therefore can be considered a valid candidate to be integrated within Devito.

%% file: conclusion/conclusion.tex
\chapter{Conclusions and future work}

\section{Conclusions}

This work has investigated the OPS intermediate representation as a way of allowing Devito operators to be executed on GPU architectures. In particular the contributions made are:

\begin{itemize}
    \item Implemented an OPS backend in Devito
    \begin{itemize}
        \item This provides the core features required to use OPS in Devito, including code generation, compilation and execution. This also gives Devito developers with a starting point to bring the OPS backend in production and conduct further experiments on GPUs.
    \end{itemize}
    \item Performance evaluation of the OPS backend
    \begin{itemize}
        \item This has shown significant speedups on GPUs compared to the same operators executed on CPUs with the core Devito backend. The performance observed on GPUs has been satisfactory and therefore encourages further work in this space. 
    \end{itemize}
    \item Software evaluation of OPS
    \begin{itemize}
        \item The main flaws in OPS as a piece of software have been outlined, along with possible ways of mitigating them and improving the software quality of the OPS library, in particular of the OPS translator. While these flaws are a reason for concern, they should not stop the adoption of the OPS API in Devito.
    \end{itemize}
\end{itemize}

Given what has been presented in this report, it is the opinion of the writer that the OPS intermediate representation is a valid path for Devito to target GPUs and therefore recommend pursuing further exploration in this space to fully integrate OPS in future Devito releases.

\section{Future work}

In this section we will detail some limitations to what is presented in this thesis and how these could be solved with further evaluations.

The necessary steps to bring the OPS backend in production are then detailed.

\subsection{Further evaluation}

While the results presented in this report are fairly indicative of what performance could be expected in applications targeted by Devito, benchmarks for common seismic operators have not been performed due to time constraints that did not allow the required features to be developed.

Currently, the OPS backend is not able to run these operators correctly. This is due to them having parts of the computation within the time loop not offloaded to OPS. In some cases (for example, computation for sources and receivers in the acoustic wave equation problem) this is not actually possible and therefore would require additional features to manipulate OPS owned data as needed.

Implementing these feature would allow more experimental work to be performed and have a clearer picture of what the performance on GPUs would be for the problem class targeted by Devito.

\subsection{Adoption}

The code developed as part of this project is experimental and has not gone through the Devito development cycle to be merged in the \texttt{master} branch for use in production. However, it provides a solid base from where to start integration of OPS in the next release of Devito.

This would include:
\begin{itemize}
    \item Rewriting some parts of the code to better align with the Devito design
    \item Splitting the already developed backend in smaller individual features to allow easier code reviews
    \item Write a comprehensive testing suite for the new features provided by the OPS backend
    \item Develop missing features as outlined in the previous section
\end{itemize}

This should bring Devito to a stage where it can target GPUs effectively with the added bonus of not having to directly worry about performance optimizations (apart from, possibly, very specific examples) for GPUs as these will be offloaded to OPS,